\documentclass[12pt,a4paper]{article}
\pdfoutput=1
\usepackage[format=hang]{caption}
\usepackage{epsfig,amssymb,amsmath,graphicx,subcaption,verbatim,hyperref,xcolor,ulem,epstopdf,psfrag,pstool,braket,array,enumerate,multirow,wrapfig}
\usepackage{graphics,color}

\numberwithin{equation}{section}

\begin{document} 
	
\title{\vskip -70pt
	\begin{flushright}
		{\normalsize DAMTP-2015-67} \\
	\end{flushright}
	\vskip 60pt
	{\bf {\LARGE Vibrational quantisation of the $B=7$ Skyrmion}}\\[20pt]}
\author{\bf {\Large C. J. Halcrow\footnote{C.J.Halcrow@damtp.cam.ac.uk}  } \\[25pt]
	Department of Applied Mathematics and Theoretical Physics\\
	University of Cambridge\\
	Wilberforce Road, Cambridge CB3 0WA, UK}

\date{October 2015}
\maketitle
\vskip 40pt

\begin{abstract}
We consider the inclusion of the most important vibrational modes in the quantisation of the dodecahedral $B=7$ Skyrmion. In contrast to a rigid body quantisation, this formalism allows a spin $\frac{3}{2}$ state to lie below the spin $\frac{7}{2}$ state, in agreement with experimental data. There is also a low lying spin $\frac{1}{2}$ state and two spin $\frac{5}{2} $ states. We find that the excited spin $\frac{7}{2}$ state has a smaller root mean square charge radius than the other states. This prediction is an important signature of the Skyrme model, in conflict with more conventional nuclear models.
\end{abstract}

\vskip 80pt

\newpage
	
\section{Introduction}
	
The Skyrme model is a non-linear field theory of pions which admits soliton solutions called Skyrmions \cite{Sk}. These are classically stable due to the topology of the system and each Skyrmion has a conserved topological charge, $B$. After quantisation Skyrmions are identified as nuclei with topological charge equal to baryon number.
	
The theory is non-renormalisable and so a first principles quantisation is beyond current methods. Instead, one must reduce the degrees of freedom in the problem to a finite number and quantise these. Each charge $B$ Skyrmion may be separated into $B$ charge one Skyrmions. These have six zero modes, three rotations and three translations. Thus to calculate quantities such as the binding energy of a nucleus one should take account of at least $6B$ degrees of freedom. Unfortunately this means quantising on a $6B$ dimensional space and little progress has been made, even for $B=2$ \cite{AttDeu}. Instead, one must select a subset of modes.
	
The simplest idea is to only include the zero modes of the Skyrmion, those transformations which leave the static energy unchanged. These are the rotations and isorotations (we stay in the centre of mass frame, allowing us to ignore translations). This procedure ignores vibrational modes, dynamical oscillations around the Skyrmion. Zero mode quantisation has had some success, such as reproducing the energy spectra of some light nuclei \cite{Light} and a natural description of the Hoyle state \cite{Hoyle}. However, there are also some failures. For example, the binding energies are all much too large. This is to be expected when we truncate the degrees of freedom from $6B$ to $6$.
	
Another failure of zero mode quantisation is the prediction of a spin $\frac{7}{2}$ ground state for the $^7$Be/$^7$Li isodoublet. The dodecahedral symmetry of the $B=7$ Skyrmion rules out low energy states with spin $\frac{1}{2}$, $\frac{3}{2}$ and $\frac{5}{2}$. In reality, experimental data show that all these states exist and the ground state has spin $\frac{3}{2}$. The first excited state of $^7$Li has spin $\frac{1}{2}$ and lies $0.5$ MeV above the ground state while the spin $\frac{7}{2}$ state is the second excited state lying $4.6$ MeV above. In this paper we shall see that the inclusion of vibrational modes in the quantisation procedure resolves this problem.

The $^7$Li and $^7$Be nuclei are special. Among all nuclei with $B<30$ they are the only ones that have an observed spin $\frac{7}{2}$ state lying below the lowest spin $\frac{5}{2}$ state. The $B=7$ Skyrmion is also special. It has the largest finite symmetry group of any known Skyrmion with non-zero pion mass. We shall see that this large symmetry group is the reason why the spin $\frac{7}{2}$ state has abnormally low energy.

The $^7$Li nucleus is usually described using a cluster model \cite{Tang} which asserts that the nucleus is made of two interacting clusters. These are an alpha particle and a tritium nucleus. This model successfully reproduces the energy spectrum and some electrostatic properties of the nucleus. We shall see that the inclusion of vibrational modes in Skyrmion quantisation highlights a connection between the Skyrme model and the ideas of clustering.

This paper is organised as follows. In section $2$ we review the Skyrme model and the structure of the $B=7$ vibrational space. We discuss how one should include vibrations in the quantisation procedure and the effects of the Finkelstein-Rubinstein constraints in section $3$. Details of the quantisation are laid out in section $4$, alongside the results of our calculations and a comparison with the experimental data.
	
\section{The $B=7$ Skyrmion and its vibrational space}
	
\subsection{The Skyrme Model}
	
The Skyrme model can be defined in terms of the three pion fields,  $ \boldsymbol\pi(t,\boldsymbol{x})$. These are combined into an $SU(2)$-valued field
\begin{equation}
U(t,\boldsymbol{x}) = \sigma(t,\boldsymbol{x}) + i \boldsymbol{\pi}(t,\boldsymbol{x})\cdot \boldsymbol\tau \, ,
\end{equation}
where $\boldsymbol\tau$ are the Pauli matrices and $\sigma$ is an auxiliary field which satisfies $\sigma^2 + \boldsymbol\pi\cdot\boldsymbol\pi = 1$. This ensures that $U \in SU(2)$. Many quantities are most easily expressed in terms of the right current $R_\mu = (\partial_\mu U)U^\dagger$. The Lagrange density is given by
\begin{equation}
\mathcal{L} = -\frac{F_\pi^2}{16}\text{Tr}\left( R_\mu R^\mu\right) + \frac{1}{32e^2}\text{Tr}\left([R_\mu,R_\nu][R^\mu,R^\nu]\right) + \frac{1}{8}m_\pi^2F_\pi^2\text{ Tr}(U-\boldsymbol{1}_2)
\end{equation}
where $F_\pi$ is the pion decay constant, $e$ is a dimensionless parameter and $m_\pi$ is the pion mass. It is more natural to work in Skyrme units. In these, the energy and length units are $F_\pi / 4e$ and $2/eF_\pi$ respectively. The Lagrangian becomes
\begin{equation}
L = \int -\frac{1}{2}\text{Tr}\left( R_\mu R^\mu\right) + \frac{1}{16}\text{Tr}\left([R_\mu,R_\nu][R^\mu,R^\nu]\right) + m^2\text{ Tr}(U-\boldsymbol{1}_2)\, d^3x
\end{equation}
where $m=2m_\pi/e F_\pi$ is the dimensionless pion mass. 
	
A Skyrmion is a solution of the field equations which minimises the static energy. This is interpreted as the classical mass of the Skyrmion and is given by
\begin{equation} \label{massf}
\mathcal{M}_B = \int -\frac{1}{2}\text{Tr}\left( R_i R_i\right) - \frac{1}{16}\text{Tr}\left([R_i,R_j][R_i,R_j]\right) - m^2\text{ Tr}(U-\boldsymbol{1}_2)\, d^3x \, .
\end{equation}
For this to be finite the Skyrme field must take a constant value, $U = \boldsymbol{1}_2$, at spatial infinity. This one point compactification of space means that $U$ is a map from $\mathbb{R}^3 \cup  \{\infty\} \cong S^3$ to $SU(2)$, which is topologically equivalent to $S^3$. These maps are labelled by an integer as $\pi_3(S^3) = \mathbb{Z}$. The integer is identified with the baryon number, $B$, and can be calculated explicitly from the Skyrme field,
\begin{equation}
B = \int \mathcal{B}(x)\, d^3x=-\frac{1}{24\pi^2}\int \epsilon_{ijk}\text{Tr}(R_i R_j R_k)\, d^3x \, 
\end{equation}
where $\mathcal{B}$ is the baryon density.
	
To visualise a Skyrmion we plot a surface of constant baryon density. This is then coloured to express the direction of the pion field, $\hat{\boldsymbol\pi}$, as it varies over the surface. We use the same colouring scheme as in \cite{108}. The Skyrmion is coloured white/black when $\hat{\pi}_3$ equals $\pm 1$ and red, green and blue when $\hat{\pi}_1+i\hat{\pi}_2$ is equal to $1$, $\exp(2\pi i /3)$ and $\exp(4\pi i /3)$ respectively.
	
\subsection{The vibrational space of the $B=7$ Skyrmion}
	
\begin{figure}
	\centering
	\includegraphics[width=1.6in]{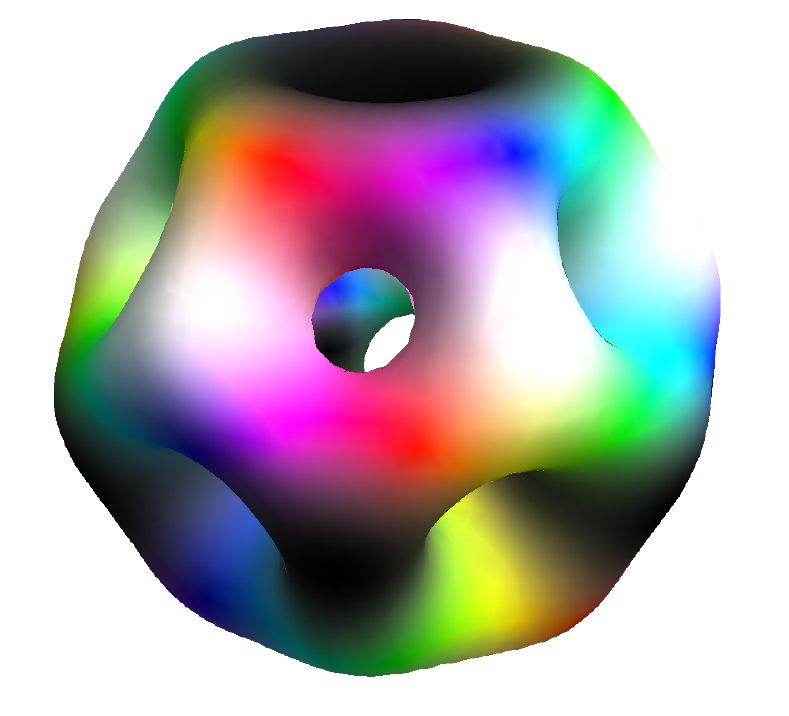}
	\caption{A surface of constant baryon density for the $B=7$ Skyrmion.}
	\label{fig:B7Sky}
\end{figure}

The $B=7$ Skyrmion has dodecahedral symmetry as seen in figure \ref{fig:B7Sky}. There is $D_5$ symmetry around each face of the Skyrmion and $D_3$ symmetry around each vertex. These, alongside the additional reflection symmetry, generate the full symmetry group of the Skyrmion $Y_h$.
	
The vibrational space was numerically generated and studied in \cite{Kim} by considering small perturbations around the $B=7$ Skyrmion. Two low frequency modes were found, one of which had a clear physical interpretation and a clean peak in the power spectrum. We will assume that this is the lowest energy vibrational mode and exclude all others from our analysis. Each point in the vibrational space corresponds to a deformed Skyrme configuration. Our aim is to understand the structure of the space and to find subspaces where the Skyrme configurations have enhanced symmetry.
	
The vibrational mode we consider has five fold degeneracy and so spans a $5$-dimensional vibrational space which we denote $\mathcal{V}_5$. Each point $\boldsymbol v \in \mathcal{V}_5$ corresponds to a quadrupole deformation tensor of the Skyrmion, $Q(\boldsymbol{v})$. There is a natural mapping from a hyperplane in $\mathbb{R}^6$ (isomorphic to $\mathcal{V}_5$) to the space of quadrupole tensors. It is
\begin{equation} \label{r6toquad}
(v_1,v_2,v_3,v_4,v_5,v_6) \mapsto \begin{pmatrix}
v_1 & 2^{-\frac{1}{2}} v_6 & 2^{-\frac{1}{2}} v_5 \\
2^{-\frac{1}{2}} v_6 & v_2 & 2^{-\frac{1}{2}} v_4 \\
2^{-\frac{1}{2}} v_5 & 2^{-\frac{1}{2}} v_4 & v_3
\end{pmatrix} \, ,
\end{equation}
where $\boldsymbol v$ satisfies $(1,1,1,0,0,0)\cdot \boldsymbol v = 0$ to ensure the quadrupole tensor is traceless. We can add vectors on the hyperplane; this is equivalent to adding the quadrupole tensors in $\mathbb{R}^3$. We choose the normalisation so that a unit vector $\boldsymbol{\hat{v}}$ maps to a quadrupole which satisfies $Q_{ij}Q_{ij} = 1$. Each quadrupole tensor, $Q(\boldsymbol v)$, has an associated symmetry group which acts on $\mathbb{R}^3$. Any symmetry shared by the quadrupole tensor and the $B=7$ Skyrmion is a symmetry of the Skyrme configuration at the point $\boldsymbol v$.

\begin{figure}
	\centering
	\includegraphics[width=5in]{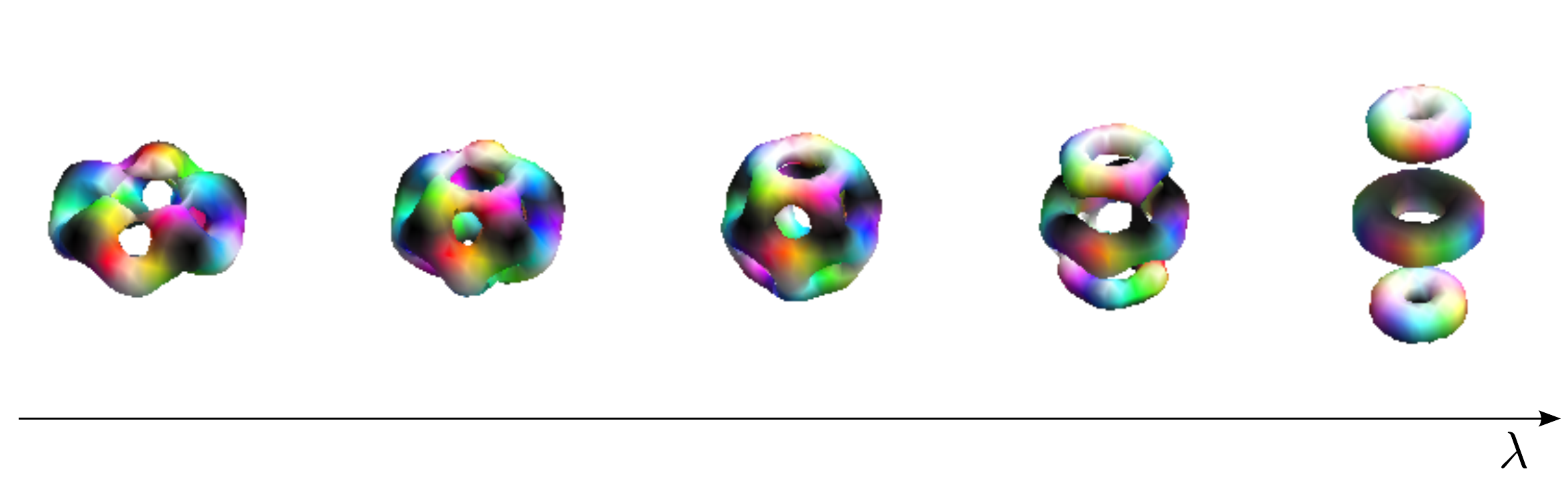}
	\caption{A vibration in $\mathcal{V}_5$ which preserves $D_5$ symmetry. The parameter $\lambda$ measures the amplitude of the vibration. This figure was generated using the gradient flow approximation to dynamics. The minimum energy Skyrmion is at $\lambda=0$. This deforms into a ring-like configuration for $\lambda<0$ and three clusters for $\lambda>0$.}
	\label{fig:D5vib}
\end{figure}

In \cite{Kim} it was found that the vibration we consider preserves the Skyrmion's $D_5$ symmetry along certain lines in $\mathcal{V}_5$. Physically, this vibration pulls on two opposite faces of the dodecahedron and breaks the Skyrmion into three clusters: a $B=3$ torus sandwiched between two $B=2$ tori. This can happen in six ways as there are six pairs of faces on the Skyrmion. Hence there are six special lines in $\mathcal{V}_5$ which preserve $D_5$ symmetry. They are evenly spaced and are aligned with the vertices of a regular $5$-simplex. We must position the $5$-simplex in $\mathcal{V}_5$ so that each vertex, $\boldsymbol{v}_a$, maps to a quadrupole tensor which is circle invariant around the axis passing through the Skyrmion faces that are being pulled upon. This ensures that the Skyrme configuration at $\boldsymbol{v}_a$ preserves $D_5$ symmetry. We use the Veronese mapping to help us. This is a map from $\mathbb{R}P^2$ to a $2$-dimensional subspace of $\mathcal{V}_5$. Explicitly it takes
\begin{equation}
(x_1,x_2,x_3) \mapsto \left(x_1^2 - \frac{1}{3}r^2,x_2^2 - \frac{1}{3}r^2,x_3^2 - \small\frac{1}{3}r^2,x_2 x_3,x_1x_3,x_1 x_2\right)\, .
\end{equation}
This then maps to a quadrupole via \eqref{r6toquad} which is circle invariant around $(x_1,x_2,x_3)$. For example, the Skyrmion has $D_5$ symmetry around the axis $\boldsymbol{x}_1=(0,0,1)$. This goes, via the Veronese mapping, to the $6$-vector
\begin{equation}
\boldsymbol{v}_1 = (-6^{-\frac{1}{2}},-6^{-\frac{1}{2}},(2/3)^\frac{1}{2},0,0,0)
\end{equation}
which maps to the quadrupole
\begin{equation}
Q_1 = \begin{pmatrix}
-6^{-\frac{1}{2}} & 0 & 0 \\
0 & -6^{-\frac{1}{2}} & 0 \\
0 & 0 & (2/3)^\frac{1}{2}
\end{pmatrix} \, .
\end{equation}
This is circle invariant around $\boldsymbol{x}_1$ as desired. Repeating this process, we may generate the vertices of the $5$-simplex in $\mathcal{V}_5$ from the lines which pass through the faces of dodecahedron. This procedure has the corollary that all six vertices of the $5$-simplex lie on the $2$-dimensional Veronese surface. We denote the $5$-simplex vertices as $\boldsymbol{v}_a \in \mathcal{V}_5$ and the corresponding quadrupole tensors $Q_a$; these are circle invariant around $\boldsymbol{x}_a$. Any configuration which lies on the line $\lambda \boldsymbol{v}_a \in \mathcal{V}_5,\, \lambda \in \mathbb{R}$ has $D_5$ symmetry. The parameter $\lambda$ is the amplitude of the vibration. For $\lambda > 0$ the Skyrmion deforms as described above: a pair of opposite faces are pulled upon. When $\lambda < 0$ the faces are pushed together and the Skyrmion flattens out. The full vibration is displayed in figure \ref{fig:D5vib}.

We may use the geometry of the $5$-simplex to find additional symmetric subspaces in $\mathcal{V}_5$. The planes passing through an edge of the simplex can be written as
\begin{equation}
\mu \boldsymbol{v}_a + \nu \boldsymbol{v}_b\, ,
\end{equation}
where $a \neq b$ and $\mu, \nu \in \mathbb{R}$. The corresponding quadrupole has a $C_2$ symmetry, shared with the $B=7$ Skyrmion, about the axis $\boldsymbol{x}_a \times \boldsymbol{x}_b$. This is enhanced to a $D_2$ symmetry when $\mu=\nu$.

\begin{figure}
	\centering
	\includegraphics[width=1.6in]{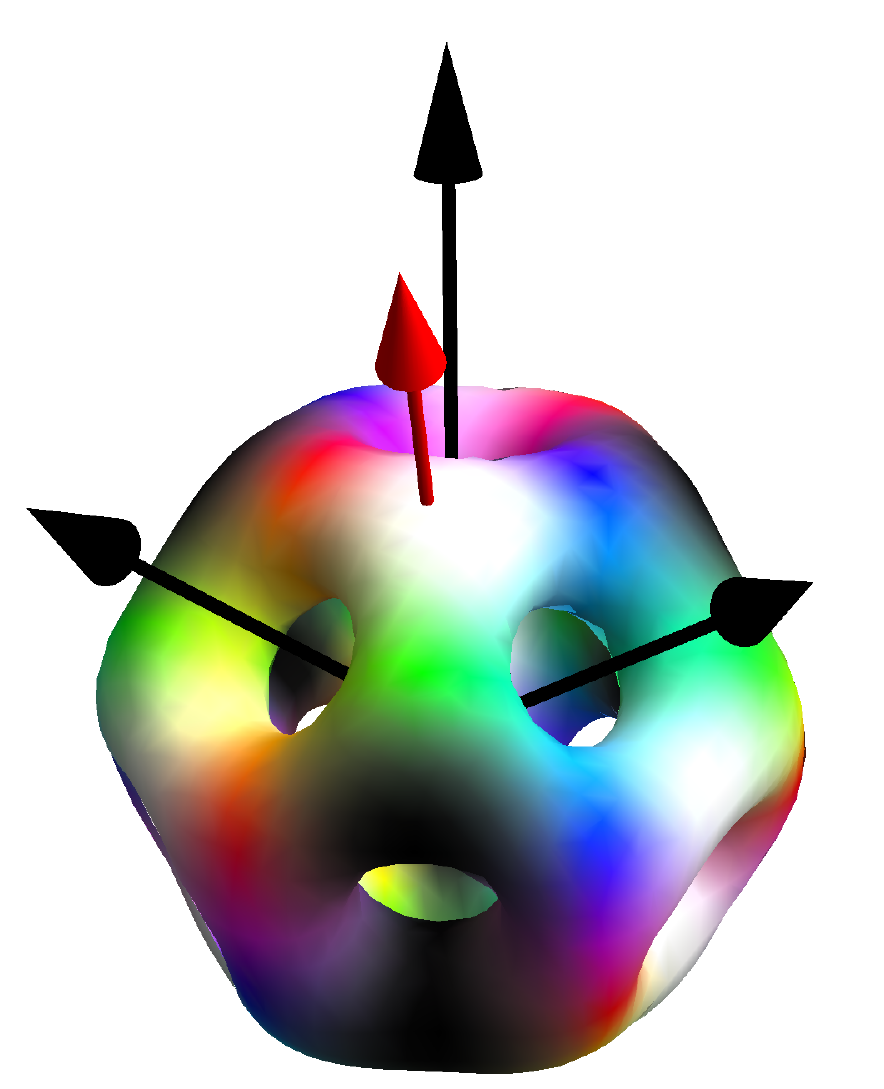}
	\caption{When three faces of the Skyrmion are pulled equally, a $D_3$ symmetry remains. The sum of the quadrupoles which pull on the faces of the Skyrmion give a quadrupole which is circle invariant about the red axis which passes through a vertex as shown.}
	\label{fig:3faces}
\end{figure}

The $5$-simplex has $20$ triangular faces. A line passing through the centre of a face takes the form
\begin{equation} \label{3lines}
\lambda(\boldsymbol{v}_a+\boldsymbol{v}_b+\boldsymbol{v}_c)\, ,
\end{equation}
where $a \neq b \neq c$. In fact, this line passes through two triangular faces which are dual to each other. Thus there are only ten distinct lines. The quadrupole tensor derived from \eqref{3lines} has only two distinct eigenvalues. Thus it is circle invariant around the eigenvector of the non-degenerate eigenvalue. This eigenvector passes through a vertex of the $B=7$ Skyrmion which has $D_3$ symmetry. Thus the Skyrme configurations on these $10$ lines in $\mathcal{V}_5$ retain $D_3$ symmetry. Note that, since these quadrupoles are circle invariant, these points in $\mathcal{V}_5$ also lie on the Veronese surface discussed earlier. It is instructive to view the physical picture. When $\lambda > 0$ the three component quadrupole tensors pull on three pairs of opposite faces. Three faces always surround a vertex of the Skyrmion, as do the opposite faces; the remaining three pairs form a ring around its centre. The quadrupole tensors around the vertex sum to give a quadrupole which pulls in the direction of the surrounded vertex. This is seen in Figure \ref{fig:3faces}. When large, this vibration breaks the Skyrmion into two $B=3$ Skyrmions sandwiching a $B=1$ Skyrmion. When $\lambda < 0$ the faces surrounding the vertex are pushed upon and the $B=7$ Skyrmion breaks into $7$ individual $B=1$ Skyrmions. 

\begin{figure}[b!]
	\centering
	\includegraphics[width=5in]{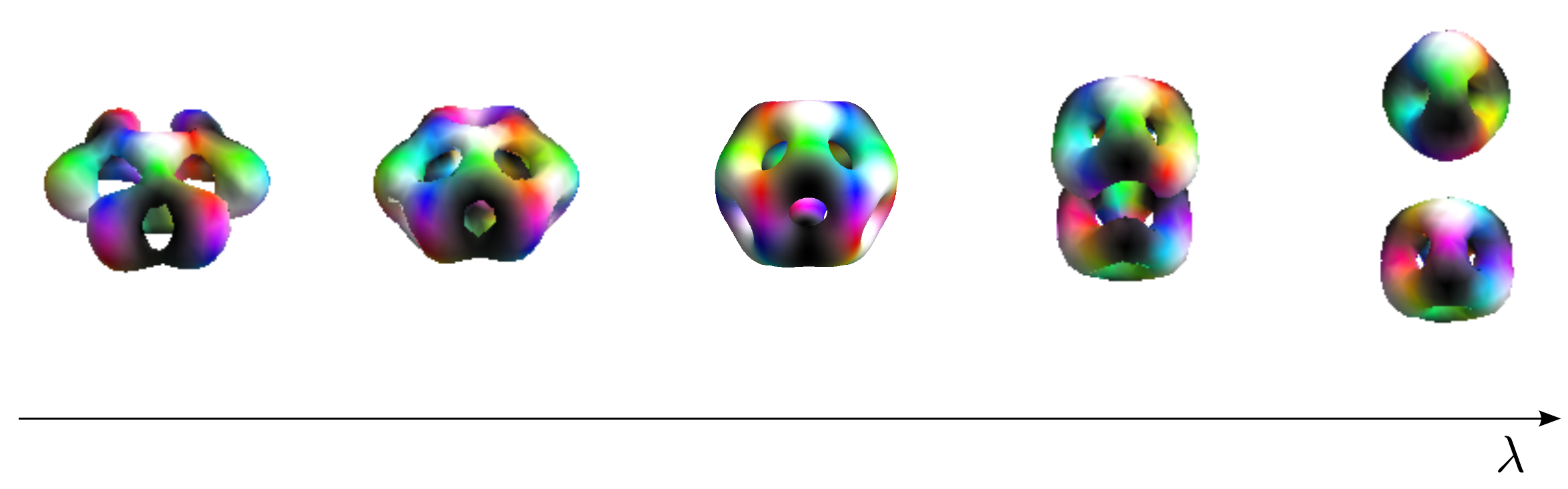}
	\caption{A vibration in $\mathcal{V}_5$ which preserves $C_3$ symmetry. The parameter $\lambda$ measures the amplitude of the vibration. The minimum energy Skyrmion is at $\lambda=0$. This deforms into seven individual distorted Skyrmions for $\lambda<0$ and two clusters for $\lambda>0$.}
	\label{fig:C3vib}
\end{figure}

The analysis so far is based on small perturbations around the Skyrmion. We believe that there will be a bifurcation where the exact symmetries discussed above will break. For example, the $D_3$ symmetry will break to a $C_3$ symmetry. This allows the asymptotic configuration in $\mathcal{V}_5$ to be a two-cluster system consisting of a $B=3$ and $B=4$ Skyrmion. This has lower energy than the three-cluster system described in the previous paragraph. These are hard to distinguish near the origin of $\mathcal{V}_5$ and so the difference will not be apparent in our analysis. Thus we shall assume that the Skyrme configuration along this vibration, at large amplitudes, will be the $C_3$ symmetric $3+4$ cluster configuration instead of the $D_3$ symmetric $3+1+3$ configuration. The entire vibration is displayed in figure \ref{fig:C3vib}.

\section{Quantisation on the vibrational space}
	
Our aim is to quantise the $B=7$ Skyrmion taking the lowest energy vibrational mode into account. The manifold we quantise must contain all configurations in $\mathcal{V}_5$ including those generated by rotations and isorotations. Explicitly the manifold is
\begin{equation}
\mathcal{N} = \frac{\mathcal{V}_5 \times SU(2) \times SU(2)}{D}
\end{equation}
where $D$ is a finite group encoding the dodecahedral symmetry of the Skyrmion. We can think of this manifold as a family of Skyrme configurations parametrised by vibrational ($\boldsymbol s$), rotational ($\phi,\theta,\psi$) and isorotational ($\alpha, \beta, \gamma$) coordinates. The angular coordinates are two sets of Euler angles. To quantise we promote all these parameters to dynamical degrees of freedom by allowing them to depend on time. This ansatz allows us to define angular velocities $\boldsymbol{b}$, isoangular velocities $\boldsymbol{a}$ and find the kinetic energy
\begin{equation}
T = \frac{1}{2}(\dot{\boldsymbol s},\boldsymbol{a}, \boldsymbol{b}).g(\boldsymbol s).(\dot{\boldsymbol s}, \boldsymbol{a}, \boldsymbol{b})^T\, ,
\end{equation} 
where $g(\boldsymbol s)$ is the metric on $\mathcal{N}$, which depends on the Skyrme configuration at $\boldsymbol s$.
	
With the kinetic energy written in this way, the quantum kinetic operator is well known \cite{LB}. It is proportional to the Laplace-Beltrami operator, $\Delta$. Explicitly
\begin{equation}
\Delta = \frac{1}{\sqrt{|g|}}\partial_i\left(\sqrt{|g|}g^{ij}\partial_j\right) \, ,
\end{equation}
where $|g|$ is the determinant of the metric. This preserves the classical symmetries of the kinetic energy after quantisation. 

The potential energy, $V(\boldsymbol s)$, is the mass \eqref{massf} of the configuration at $\boldsymbol s$. The Hamiltonian on $\mathcal{N}$ is
\begin{equation} \label{ham}
\mathcal{H} = -\frac{\hbar^2}{2} \Delta + V(\boldsymbol s) \, .
\end{equation}
To find bound states of definite energy we solve the stationary Schr\"{o}dinger equation arising from this Hamiltonian,
\begin{equation} \label{Schro}
\mathcal{H}\Psi = E \Psi \, .
\end{equation}
Formally, the wavefunction $\Psi$ is a section of a complex line bundle over $\mathcal{N}$. 
	
There are constraints on $\Psi$ which encode the fact that nucleons are fermions. These are the Finkelstein-Rubinstein (FR) constraints \cite{FRc}. They can be written in terms of the classical symmetries of the Skyrmion. For example, the $B=7$ Skyrmion is invariant under a $2 \pi / 5$ rotation around the $3$-axis followed by a $-4 \pi / 5$ isorotation around the $3$-axis in isospace. In operator form, this $C_5$ symmetry puts the following constraint on the wavefunction
\begin{equation} \label{FR1}
e^{\frac{2\pi i}{5}\hat{L}_3}e^{-\frac{4\pi i}{3}\hat{K}_3} \Psi = -\Psi \, ,
\end{equation}
where $\hat L_3$ and $\hat K_3$ are the body fixed angular momentum operators defined in the usual way. Similarly, the $C_3$ symmetry gives the constraint
\begin{equation} \label{FR2}
e^{2\pi i /3\, \boldsymbol{n}_1 \cdot \boldsymbol{\hat L}}e^{i\xi \,\boldsymbol{n}_2 \cdot \boldsymbol{\hat K}}\Psi=\Psi \, . 
\end{equation}
where $\boldsymbol{n}_1=(-\sqrt{\frac{2}{15}(5-\sqrt{5})},0,\sqrt{\frac{1}{15}(5+2\sqrt{5})})$ is a vector which passes through one of the dodecahedron's vertices while $\xi$ and $\boldsymbol{n}_2$ define the isorotation required to return the Skyrmion to its original colouring. The FR signs can be calculated using the rational map ansatz \cite{RMan}. The procedure is set out in \cite{Steffan}. 
	
The constraints \eqref{FR1} and \eqref{FR2} both apply when the Skyrmion has dodecahedral symmetry. This occurs at the origin of $\mathcal{V}_5$. For a generic point $\boldsymbol s$ there is no symmetry and thus no constraints. In the previous Section we found lines in $\mathcal{V}_5$ which had enhanced symmetry. One set of these preserved $D_5$ symmetry. Thus, on these lines, only a constraint such as \eqref{FR1} applies, as well as an additional constraint which enhances the $C_5$ symmetry to $D_5$. Another set of lines preserved $C_3$ symmetry, meaning the wavefunction must satisfy a constraint such as \eqref{FR2} on these.
	
Now the problem is formulated. To include the lowest vibrational mode when studying the states of $^7$Li/$^7$Be  we must solve \eqref{Schro}, an $11$-dimensional Schr\"odinger equation, subject to \eqref{FR1} and \eqref{FR2} at $\boldsymbol s=0$, just a constraint such as \eqref{FR1} on six lines in the vibrational space (representing the $D_5$ preserving directions in $\mathcal{V}_5$) and a constraint such as \eqref{FR2} on ten lines. There are further constraints on the edges of the $5$-simplex. To set up and solve this problem rigorously is too hard and so we will make some simplifying assumptions below.
	
In certain cases the metric, $g$, will simplify due to the symmetries of the system. In particular the kinetic operator can separate into a part which only acts via the rotational coordinates and a part which only acts via the vibrational ones. We denote this as
\begin{equation}
\Delta = \Delta_s + \nabla^2 \, .
\end{equation}
We can then solve the Schr\"{o}dinger equation using separation of variables
\begin{equation} \label{SepWv}
\Psi = u(\boldsymbol s)\Theta(\phi,\theta,\psi,\alpha,\beta,\gamma) \, ,
\end{equation}
where we call $u$ the vibrational wavefunction and $\Theta$ the rotational wavefunction. The latter is a solution of the rigid body Schr\"{o}dinger equation. This problem has been studied extensively, see \cite{Light} for details. The solutions are tensor products of Wigner D-functions and each solution has six conserved quantities: total spin ($J^2$), total isospin ($I^2$), body fixed spin ($L_3$) and isospin ($K_3$) projections and space fixed spin ($J_3$) and isospin ($I_3$) projections. The space fixed projections do not affect the energy spectrum and as such they are often suppressed in the bra-ket notation where we denote the state $ D^J_{L_3 J_3}(\phi,\theta,\psi) \otimes D^I_{K_3 I_3}(\alpha,\beta,\gamma)$ as $\ket{J \, L_3} \ket{I\,K_3}$.
	
One may satisfy the FR constraints using the rotational wavefunction by taking appropriate linear combinations of the Wigner functions. Often the constraints rule out certain spin states. The dodecahedral symmetry of the $B=7$ Skyrmion rules out states with $J=\frac{1}{2}, \frac{3}{2} \text{ and } \frac{5}{2}$ and $I=\frac{1}{2}$. However, the constraints apply to the \textit{entire} wavefunction $\Psi$, not just the rotational part $\Theta$. We may alternatively satisfy the FR constraints at the origin of $\mathcal{V}_5$ by insisting that $\Psi$ vanishes there. So there are two ways to satisfy the FR constraints at $\boldsymbol s=0$:
\begin{enumerate}[(a)]
	\item The rotational wavefunction, $\Theta$, is permitted by both the FR constraints. There are no restrictions on the vibrational wavefunction, $u(\boldsymbol s)$.
	\item The vibrational wavefunction is zero at the origin, $u(\boldsymbol 0) = 0$. There are fewer restrictions on $\Theta$.
\end{enumerate}
These two options also apply on any subspace of vibrational space with enhanced symmetry.
\begin{comment}	
Let us fix $I = \frac{1}{2}$ and look for low energy states ($J \leq \frac{7}{2}$). We will now review which spin states are allowed by each symmetry we've encountered, many of these were discovered and discussed in \cite{Olga7}.
	
\begin{description}	
\item[The dodecahedron at the origin of $\mathcal{V}_5$. ] The only allowed state has $J=\frac{7}{2}$.
\item[The $D_5$ symmetric lines on the vertices of the $5$-simplex. ] No spin $\frac{1}{2}$ states are allowed. One spin $\frac{3}{2}$ state is allowed and one spin $\frac{5}{2}$ states is allowed. There are two permitted $\frac{7}{2}$ states: the dodecahedral state and an orthogonal one.
\item[The $C_2$ symmetry present on the edges of the $5$-simplex. ] This is not a very restrictive symmetry. There are two, four, six and eight states allowed with spin $\frac{1}{2}$, $\frac{3}{2}$, $\frac{5}{2}$ and $\frac{7}{2}$ respectively.
\item[The $D_2$ symmetry present at the centre of the edges of the $5$-simplex. ] There are one, two, three and four states allowed with spin $\frac{1}{2}$, $\frac{3}{2}$, $\frac{5}{2}$ and $\frac{7}{2}$ respectively.
\item[The $C_3$ symmetric lines going through the triangular faces of the $5$-simplex. ] Many states are allowed here. Two spin $\frac{1}{2}$ states, two spin $\frac{3}{2}$ states, three spin $\frac{5}{2}$ states and three spin $\frac{7}{2}$ states.
\end{description}
\end{comment}

We now look for the low energy states. A rotational wavefunction with spin $\frac{7}{2}$ is allowed everywhere in vibrational space. Thus the corresponding wavefunction, $\Psi_{J=\frac{7}{2}}$, can be of type (a). There are no spin $\frac{3}{2}$ states allowed at the origin of $\mathcal{V}_5$. Thus the spin $\frac{3}{2}$ wavefunction is of type (b) and $u_{J=\frac{3}{2}}$ must vanish at $\boldsymbol s=0$. We can schematically calculate the energy difference of these states using a harmonic approximation. The spin $\frac{7}{2}$ vibrational wavefunction is in the ground state of $\mathcal{V}_5$ while the spin $\frac{3}{2}$ wavefunction must be excited in \textit{one} direction so that it vanishes at the origin. So the spin $\frac{3}{2}$ state has one unit of vibrational energy more than the spin $\frac{7}{2}$ state. However it will have less rotational energy as the spin is smaller. The ordering of these states depends on the relative energy contributions from vibrations and rotations. We make the approximation that all other vibrations contribute equally to the states. Thus to compare these low lying states we only need to account for the vibration in one direction in $\mathcal{V}_5$, the smallest energy direction.

In the harmonic approximation the direction of the vibration does not matter as the potential on $\mathcal{V}_5$ is isotropic. However in the full model the direction will be important. A generic direction in $\mathcal{V}_5$ will break the $B=7$ Skyrmion into seven $B=1$ Skyrmions. This has high potential energy compared to the break up into clusters we saw in Section $2$ where the Skyrmion could break into fewer, higher-charge Skyrmions. Thus we believe that the smallest vibration will not be in a generic direction. Instead it will be along one of the high symmetry directions previously discussed.
	
\section{Quantising the $B=7$ Skyrmion}
	
We will now quantise the $B=7$ Skyrmion taking a single vibrational mode into account. To do this we must decide on the direction of the vibration in $\mathcal{V}_5$. We shall assume the lowest energy direction is along either a $C_3$ preserving line or a $D_5$ preserving line since these have low energy configurations asymptotically.
	
The symmetry present on these lines restricts the form of the metric which is 7-dimensional. It is standard convention to split the metric into submatrices. We follow \cite{Duet} and write
\begin{equation}
g = \begin{pmatrix}
\Lambda & \multicolumn{2}{c}{$0$} \\
\multirow{2}{*}{$0$} & U & -W \\
& -W^T & V
\end{pmatrix}
\end{equation}
where $U$, $W$ and $V$ are $3\times3$ matrices and $\Lambda$ is a scalar. The kinetic energy is invariant under the action of the symmetry group of the vibration. This restriction means that, along the symmetric lines in $\mathcal{V}_5$, the cross terms in the metric vanish and the kinetic energy is separable in the sense described in the previous Section. As such the wavefunction takes the form \eqref{SepWv} on these lines, with the vibrational parameter $s$ now 1-dimensional.
	
Consider a rotational state with spin $J$ and denote the rotational energy contribution $E_J(s)$ so that
\begin{equation}
\nabla^2 \Theta_J = E_J(s)\Theta_J \, .
\end{equation}
Note that the rotational energy contribution is a function of $s$ through its dependence on the moments of inertia which vary as the Skyrmion deforms. Then the Schr\"{o}dinger equation \eqref{Schro} reduces to the $1$-dimensional equation
\begin{equation} \label{Schro2}
\left( - \frac{\hbar^2}{2\sqrt{|g|}}\partial_s\left( \frac{\sqrt{|g|}}{\Lambda}\partial_s\right) +V(s)+E_J(s)\right)u(s) = Eu(s) \, .
\end{equation}
To solve this we must first generate $g(s)$, $V(s)$ and $E_J(s)$. We will do this using gradient flow.
	
Gradient flow generates a path of steepest descent in field space. We use the separated Skyrmion clusters as initial configurations which are then evolved in gradient flow time $\tau$ according to
\begin{equation} \label{flow}
\frac{d \boldsymbol{\pi} }{d \tau} = - \frac{\delta \mathcal{M}_7}{\delta \boldsymbol{\pi}} \, ,
\end{equation}
where $\boldsymbol{\pi}$ are the pion fields and $\mathcal{M}_7$ is the potential energy \eqref{massf}. This flow reduces the potential energy of the system and ends at a stationary point of field space. The fields $\boldsymbol{\pi}(\tau)$ approximate the Skyrme configurations along a half line in $\mathcal{V}_5$. The solution of \eqref{flow} is beyond analytic calculation and so we must use a numerical code to calculate the flow. The energy $V(\tau)$ and the metric $g(\tau)$ are calculated at numerous points during the process.
	
The metric at time $\tau$ can be expressed in terms of the currents $R_i = (\partial_i U)U^{-1}$ and $T_i = \frac{i}{2}[\tau_i,U]U^{-1}$. The moments of inertia and $\Lambda$ are given by
\begin{align}
&\Lambda = -\int \text{Tr}\left( R_\tau R_\tau + [R_\tau,R_i][R_\tau,R_i]\right) d^3 x \\
& U_{ij} = -\int \text{Tr}\left(T_iT_j +  \frac{1}{4}[R_k,T_i][R_k,T_j]\right) d^3 x \\
& W_{ij} = \int \epsilon_{jlm}x_l\text{Tr}\left(T_iR_m + \frac{1}{4}[R_k,T_i][R_k,R_m]\right)d^3 x \\
& V_{ij} = -\int \epsilon_{ilm}\epsilon_{jnp}x_lx_n\text{Tr}\left( R_mR_p + \frac{1}{4}[R_k,R_m][R_k,R_p]\right)d^3 x \, \text{.}
\end{align}

Gradient flow time is an unnatural parameter when the Skyrmion clusters are widely separated and near the dodecahedral configuration. Thus, once we have found our quantities numerically we change variables to the geodesic distance, $s$ \cite{book}. This can be defined in terms of the vibrational kinetic energy by demanding
\begin{equation}
T_\text{vib} = \frac{1}{2}\dot s^2  = \frac{1}{2}\Lambda(\tau)\dot{\tau}^2
\end{equation}
which means that
\begin{equation}
s(\tau) = \int^\tau \sqrt{\Lambda(\tau')} \, d\tau' \, .
\end{equation}

There are several advantages to this new coordinate. First, the geodesic distance is related to the cluster separation, $r$, asymptotically. We can calculate how the moments of inertia vary with $r$ and this gives an asymptotic check of the numerics. Additionally we are able to add an analytic tail to the numerically derived potential and moments of inertia. Further, we may now calculate the harmonic frequency near the origin of $\mathcal{V}_5$ and compare it to what was calculated in \cite{Kim}. We find the frequency to be $0.34$ compared with $0.302$ as found in \cite{Kim}. These are approximately the same, showing the methods are consistent. The small difference is likely due to the different pion masses used. Finally, the new coordinate simplifies the Schr\"{o}dinger equation \eqref{Schro2}. It now reads
\begin{equation} \label{Schro3}
\left( - \frac{\hbar^2}{2} \frac{d^2}{ds^2}-\frac{\hbar^2}{4}\partial_s\log(|g|)\frac{d}{ds} +V(s)+E_J(s)\right)u(s) = Eu(s)
\end{equation}
From now on, $s$ will refer exclusively to the geodesic distance.

\subsection{The $C_3$ direction} \label{C3}

The initial configuration for the $C_3$ direction is constructed using a symmetrised product ansatz of a $B=3$ Skyrmion with a $B=4$ Skyrmion. These are orientated as in figure \ref{fig:C3vib}. The $C_3$ symmetry constrains the form of the metric. We find that $U$, $V$ and $W$ are all diagonal. Further
\begin{equation}
U_{11} = U_{22}, \, V_{11} = V_{22}, \, \text{and } \, W_{11} = W_{22} \, .
\end{equation}
We have set $\Lambda = 1$ by choosing our parameter to be the geodesic distance.
	
We now look at specific rotational wavefunctions. Although this direction in $\mathcal{V}_5$ only has $C_3$ symmetry, it has approximate $D_3$ symmetry near the origin. This means that a wavefunction disallowed by $D_3$ symmetry would have extra constraints imposed on it nearby in the full vibrational space. This would increase the energy of the state. Thus we focus on states which are allowed by $D_3$ symmetry. The rotational wavefunctions we consider are presented in table \ref{D3tab}.	
{
\setlength\extrarowheight{6pt}
	\begin{table}
		\begin{center}
			\begin{tabular}{|l | l|} 
				\hline
				Spin & FR-allowed states \\ \hline
				$J=\frac{1}{2}$ & $\ket{\Theta}_{\frac{1}{2}} = \ket{\frac{1}{2},\frac{1}{2}}\ket{\frac{1}{2},\frac{1}{2}} - \ket{\frac{1}{2}, -\frac{1}{2}}\ket{\frac{1}{2}, -\frac{1}{2}}$ \\
				$J=\frac{3}{2}$ & $\ket{\Theta}_{\frac{3}{2}} =\ket{\frac{3}{2},\frac{1}{2}}\ket{\frac{1}{2},\frac{1}{2}}+\ket{\frac{3}{2},-\frac{1}{2}}\ket{\frac{1}{2},-\frac{1}{2}}$ \\
				$J=\frac{5}{2}$ & $\ket{\Theta}^{(1)}_{\frac{5}{2}} =\ket{\frac{5}{2},\frac{1}{2}}\ket{\frac{1}{2},\frac{1}{2}}+\ket{\frac{5}{2},-\frac{1}{2}}\ket{\frac{1}{2},-\frac{1}{2}}$ \\
				&
				$\ket{\Theta}^{(2)}_{\frac{5}{2}}=\ket{\frac{5}{2},-\frac{5}{2}}\ket{\frac{1}{2},\frac{1}{2}}-\ket{\frac{5}{2},\frac{5}{2}}\ket{\frac{1}{2},-\frac{1}{2}}$ \\
				$J=\frac{7}{2}$ & $\ket{\Theta}^{(1)}_{\frac{7}{2}} =\ket{\frac{7}{2},\frac{1}{2}}\ket{\frac{1}{2},\frac{1}{2}}+\ket{\frac{7}{2},-\frac{1}{2}}\ket{\frac{1}{2},-\frac{1}{2}}$ \\
				&
				$\ket{\Theta}^{(2)}_{\frac{7}{2}}=\ket{\frac{7}{2},-\frac{5}{2}}\ket{\frac{1}{2},\frac{1}{2}}-\ket{\frac{7}{2},\frac{5}{2}}\ket{\frac{1}{2},-\frac{1}{2}}$ \\
				&
				$\ket{\Theta}^{(3)}_{\frac{7}{2}}=\ket{\frac{7}{2},\frac{7}{2}}\ket{\frac{1}{2},\frac{1}{2}}-\ket{\frac{7}{2},-\frac{7}{2}}\ket{\frac{1}{2},-\frac{1}{2}}$ \\ \hline
			\end{tabular} 
		\end{center}
		\caption{The low energy states allowed by $D_3$ symmetry.}\label{D3tab}
	\end{table}
}

Consider the spin $\frac{3}{2}$ state. The full wavefunction is of the form
\begin{equation} \label{32state}
\Ket{\Psi}_\frac{3}{2} = u_\frac{3}{2}(s)\left( \, \Ket{\frac{3}{2},\frac{1}{2}}\Ket{\frac{1}{2},\frac{1}{2}}+\Ket{\frac{3}{2},-\frac{1}{2}}\Ket{\frac{1}{2},-\frac{1}{2}} \right) \, .
\end{equation} 
The vibrational wavefunction must satisfy $u_\frac{3}{2}(0)=0$ so that the full wavefunction $\Psi_\frac{3}{2}$ is consistent with the additional FR constraint \eqref{FR1} at $s=0$. The derivative must be non-zero here or the vibrational wavefunction will be trivial everywhere. Inserting \eqref{32state} into the Schr\"odinger equation \eqref{Schro} we find that $u_\frac{3}{2}$ satisfies
\begin{align}
\Bigg( \frac{\hbar^2}{2}\Big( \frac{V_{11}}{2\left(U_{11}V_{11}-W_{11}^2\right)}+\frac{3U_{11}}{2\left(V_{11}U_{11}-W_{11}\right)} + \frac{1}{U_{33}V_{33} - W_{33}^2}  \Big(\frac{9}{4}U_{33} + \frac{1}{4}V_{33} - &\frac{3}{2}W_{33}\Big) \Big) \nonumber \\
-\frac{\hbar^2}{2}\frac{d^2}{ds^2}-\frac{\hbar^2}{4}\partial_s\log(|g|)\frac{d}{ds}+ V(s)\Bigg) u_\frac{3}{2}(s) = E u_\frac{3}{2}(s)& \, .
\end{align}
We would like to understand the contributions from rotations and vibrations separately. There is no unique way to split the energy; we choose to define the rotational energy contribution as the rigid body energy of the undeformed Skyrmion, $E_J(0)$. We may then split the energy $E$ into three parts: the classical mass of the Skyrmion $\mathcal{M}_7 = V(0)$, the contribution from the rigid rotation $E_J(0)$, and the energy contribution from the vibration $\epsilon_\text{vib}$. We write $E = \mathcal{M}_7 + E_J(0) + \epsilon_\text{vib}$ and the Schr\"odinger equation becomes
\begin{equation} \label{Schro4}
\left( -\frac{\hbar^2}{2}\frac{d^2}{ds^2}-\frac{\hbar^2}{4}\partial_s\log(|g|)\frac{d}{ds} + V_\text{eff}(s)\right)u_\frac{3}{2}(s) = \epsilon_\text{vib} u_\frac{3}{2}(s)
\end{equation}
where $V_\text{eff}(s) = V(s)-\mathcal{M}_7 + E_J(s)-E_J(0)$. Note that $V_\text{eff}$(0) = 0. Equation \eqref{Schro4} is then solved numerically using a shooting technique.

There are two spin $\frac{5}{2}$ states. These have different values of $L_3$ and thus each state has a different effective potential. Each full wavefunction is of the form
\begin{equation}
\Ket{\Psi}_\frac{5}{2} = u_\frac{5}{2}(s)\Ket{\Theta}_\frac{5}{2}
\end{equation}
with each vibrational wavefunction being zero at $s=0$, just like the spin $\frac{3}{2}$ case.

There are three spin $\frac{7}{2}$ states. We focus on the lowest energy state. This is a linear combination of the three states,
\begin{equation}
\Ket{\Psi}_\frac{7}{2} = u(s) \ket{\Theta}^{(1)}_{\frac{7}{2}} + v(s) \ket{\Theta}^{(2)}_{\frac{7}{2}} + w(s) \ket{\Theta}^{(3)}_{\frac{7}{2}} 
\end{equation}
where
\begin{equation} \label{sevenhalfscond}
u(0) = v(0) = (7/18)^{1/2} \text{ and } w(0) = \sqrt{2}/3
\end{equation}
to ensure that $\Psi_\frac{7}{2}$ satisfies the additional FR constraint \eqref{FR1} at $s=0$. This gives three uncoupled Schr\"odinger equations for $u$, $v$ and $w$. Generally these three independent equations will not produce a shared energy eigenstate as the effective potential is different for each component rotational wavefunction. However, we can obtain a shared eigenvalue by enforcing an additional boundary condition that the probability distribution is maximal at the origin. This gives
\begin{equation}
0 = \frac{1}{2}\frac{d}{ds}\biggr(|\Psi|^2\biggr)\biggr\rvert_{s=0} = \left(u\dot u+ v\dot v+ w\dot w\right)\rvert_{s=0} \, .
\end{equation}
This condition, alongside \eqref{sevenhalfscond}, produces a discrete eigenvalue spectrum.

The spin $\frac{1}{2}$ state is similar to the spin $\frac{3}{2}$ state and takes a form analogous to \eqref{32state} with the same conditions on the vibrational wavefunction. However, it  has additional constraints in the full vibrational space. It must vanish on the $D_5$ preserving lines in $\mathcal{V}_5$ due to the FR constraints. The wavefunction we construct is concentrated along a $C_3$ direction. This direction is maximally far away from the $D_5$ lines. This can be seen geometrically: the $C_3$ lines go through the centre of the $5$-simplex faces while the $D_5$ lines pass through the vertices. Thus the wavefunction we construct should already be small on the $D_5$ lines. A modification is required to make the wavefunction vanish on the $D_5$ lines which will cost energy. Thus, we expect the true spin $\frac{1}{2}$ state to have higher energy than what is calculated here.
	
\subsubsection{Calibration of the model}

Before comparing our results to experimental data we must calibrate the model. All previous calibrations are based on zero mode quantisation and as such we don't necessarily expect our choice of parameters to match previous studies. The vibrational energy contribution is of order $\hbar$ while the rotational energy contribution is of order $\hbar^2$. Thus the relative energies of the states will be sensitive to the value of $\hbar$.

\begin{figure}[ht]
	\centering
	\includegraphics[width=4.5in]{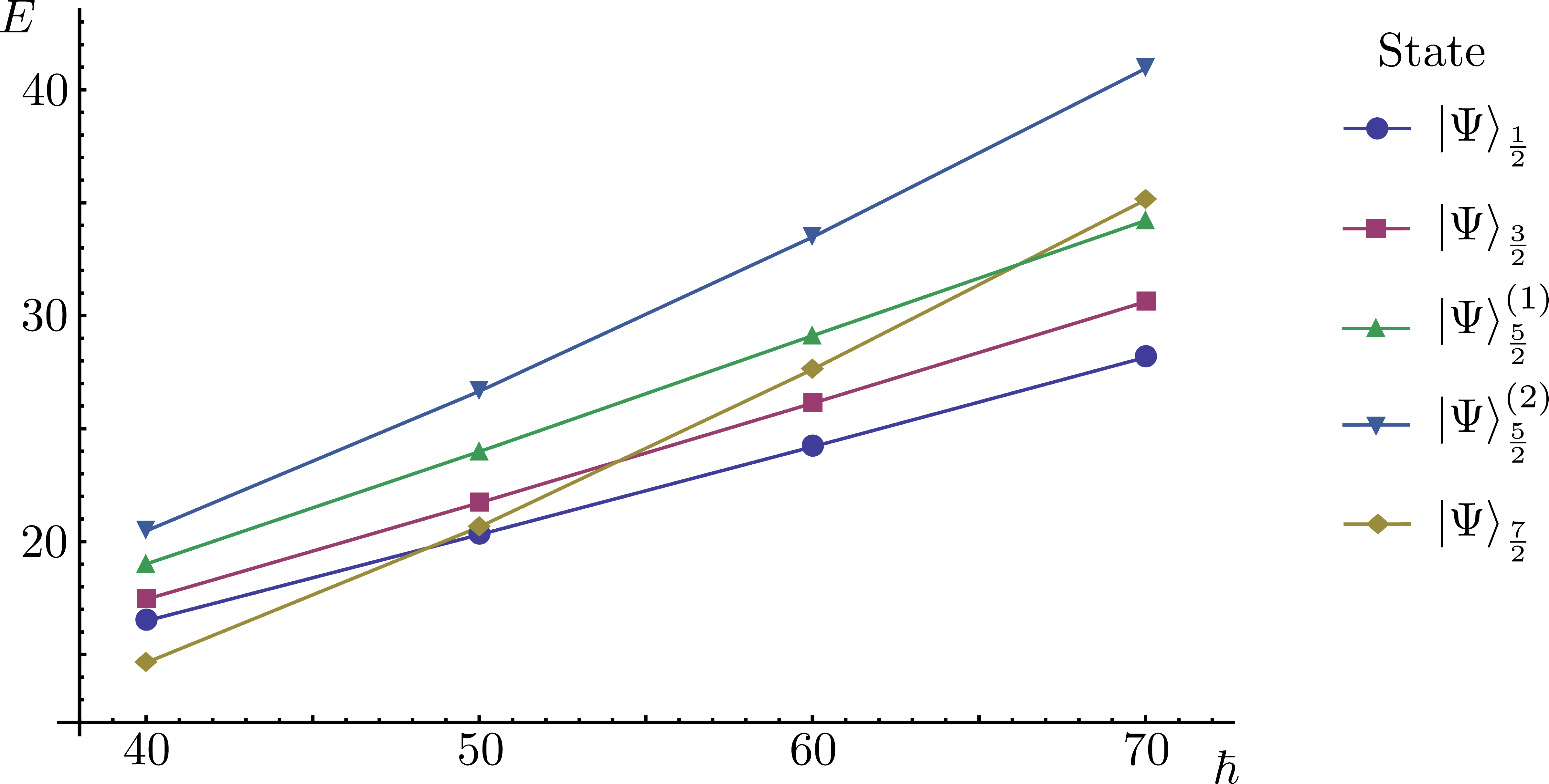}
	\caption{The quantum energy of each state (in Skyrme units) as a function of $\hbar$.}
	\label{fig:hbarvseng}
\end{figure}

In figure \ref{fig:hbarvseng} the quantum energy of each state is plotted (in Skyrme units) for various values of $\hbar$. The most important feature of the plot is that the spin $\frac{7}{2}$ state increases in energy, relative to the other states, as $\hbar$ increases. This is because the spin $\frac{7}{2}$ state has the largest rotational energy and the smallest vibrational energy; rotational effects dominate for large $\hbar$ while vibrational effects dominate for small $\hbar$. To match experimental data the spin $\frac{7}{2}$ state must lie between the spin $\frac{3}{2}$ state and the first spin $\frac{5}{2}$ state. This occurs when
\begin{equation}
55 < \hbar < 65 \, ,
\end{equation}
and as such we demand that $\hbar$ lies in this interval. For illustrative purposes we fix $\hbar=60$.

We are left to choose the value of $F_\pi$, as $\hbar$ fixes the dimensionless constant $e$ through the identity $\hbar = 2e^2$. We will consider two alternative calibrations:
\begin{itemize}
	\item[(i)] $F_\pi = 60$ MeV
	\item[(ii)] $F_\pi = 139$ MeV.
\end{itemize}
Parameter choice (i) gives a good fit to the size of the gaps in the energy spectrum but underestimates the total mass of the Skyrmion compared to the total mass of $^7$Li. Choice (ii) gives a reasonable value of the total mass but overestimates the gaps in the spectrum. We have fixed the dimensionless pion mass $m$ to $1$ throughout.
	
\subsubsection{Results}

We solved the Schr\"odinger equation \eqref{Schro4} for all states in discussed in section \ref{C3}. The numerically generated vibrational wavefunction $u(s)$, potential $V(s)-\mathcal{M}_7$ and effective potential $V_\text{eff}(s)$ for each state is plotted in table \ref{tab:C3wvs}. We also note the classical mass of the Skyrmion $\mathcal{M}_7$, the energy contribution from rotations $E_J(0)$ and the contribution from vibrations $\epsilon_\text{vib}$, as well as the total energy of each state $E$. Our results are then compared to experimental data in table \ref{Data} for each calibration (i) and (ii).

\begin{table}[ht]
	\centering
	\begin{tabular}{| l | c | l | l |} 
		\hline
		State & Vibrational wavefunction and potentials & $\mathcal{M}_7 + E_J(0) + \epsilon_\text{vib}$  & $E$ \\ \hline 
		$\ket{\Psi}_\frac{1}{2}$ & \includegraphics[width=2in]{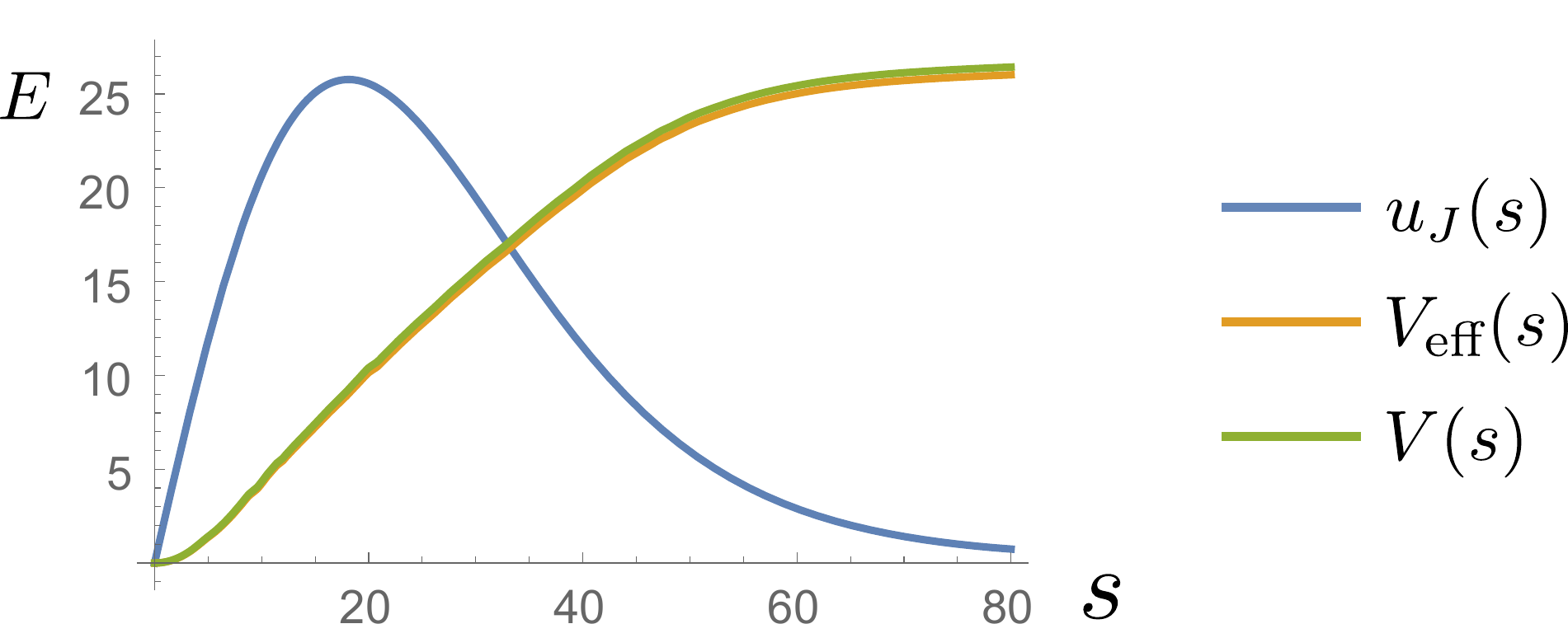} & 985.13 + 5.95 + 18.24 & 1009.32 \\ \hline		
		$\ket{\Psi}_\frac{3}{2}$ & \includegraphics[width=2in]{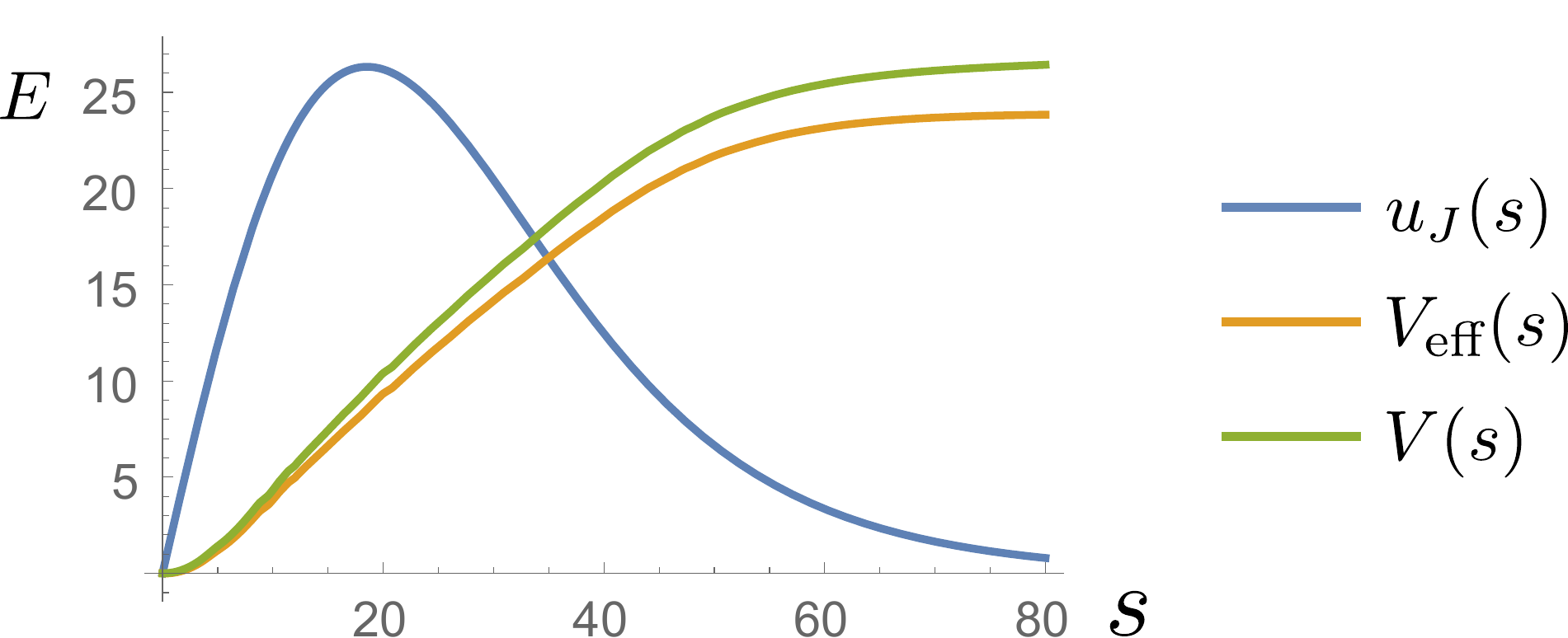} & 985.13 + 8.81 + 17.31 & 1011.25 \\ \hline
		$\ket{\Psi}_\frac{5}{2}^{(1)}$ & \includegraphics[width=2in]{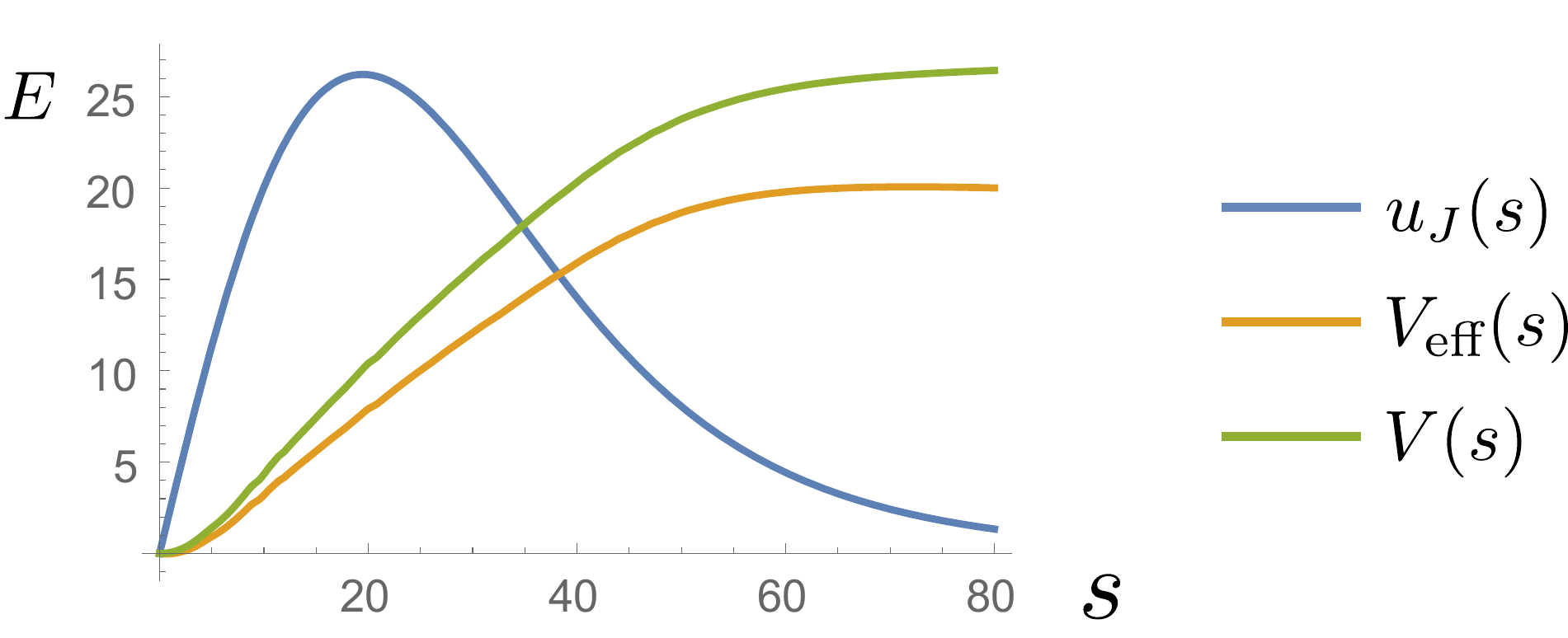} & 985.13 + 13.56 + 15.55 & 1014.24 \\ \hline
		$\ket{\Psi}_\frac{5}{2}^{(2)}$ & \includegraphics[width=2in]{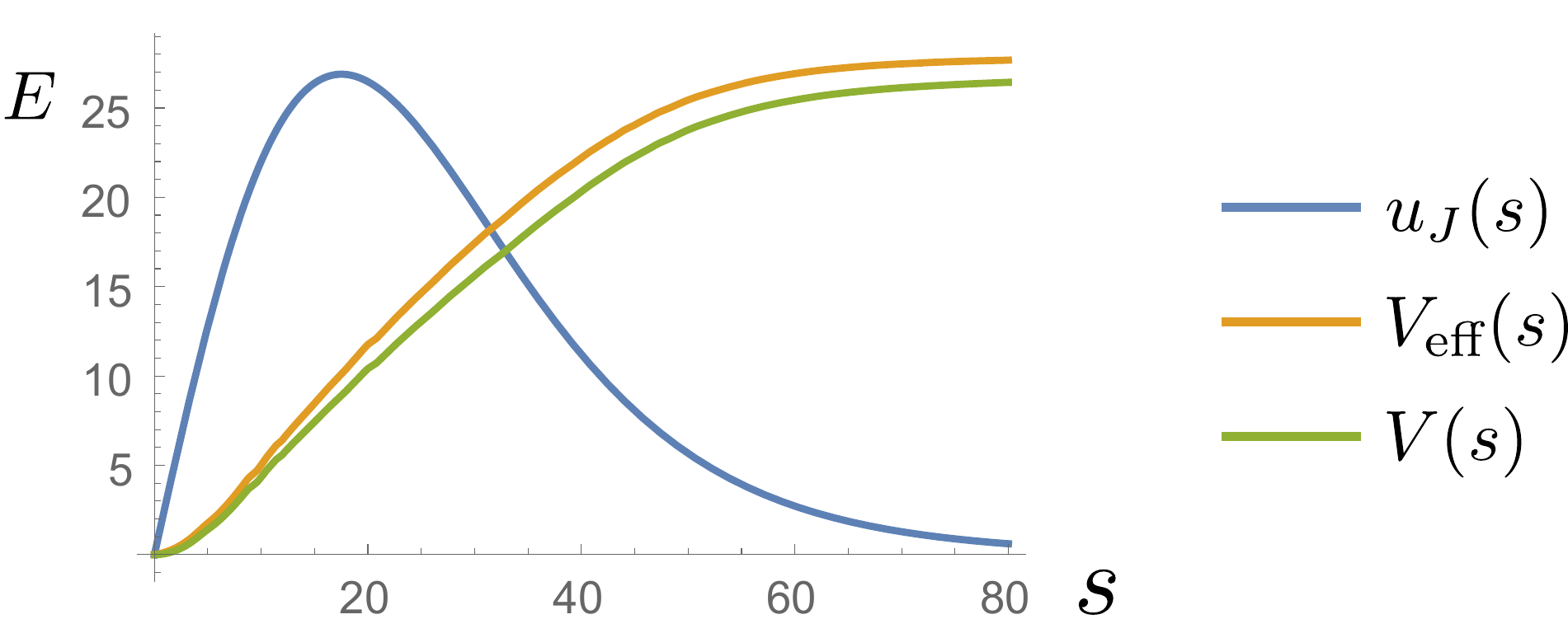} & 985.13 + 13.56 + 19.42 & 1018.11 \\ \hline
		$\ket{\Psi}_\frac{7}{2}$ & \includegraphics[width=2in]{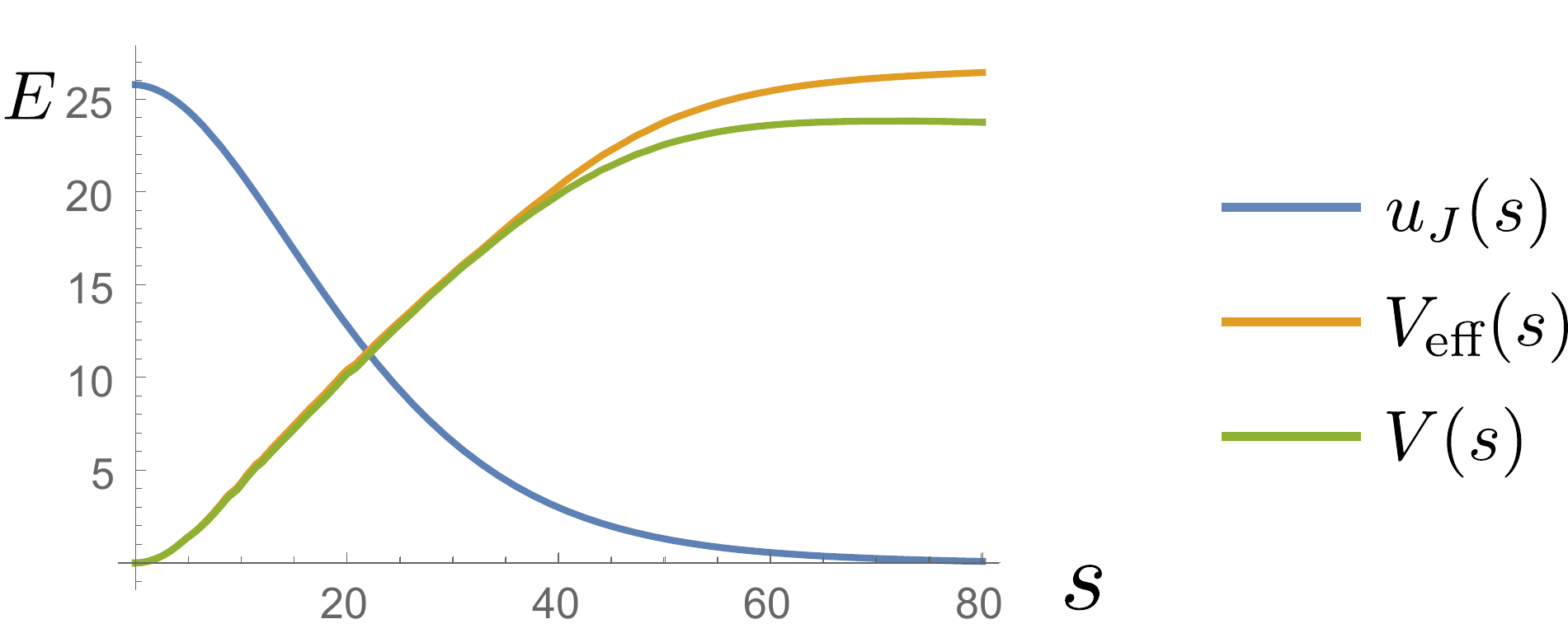} & 985.13 + 20.24 + 7.39 & 1012.76 \\ \hline
	\end{tabular} 
	\caption{The numerical results for quantisation along the $C_3$ direction. We display the vibrational wavefunction, potential and effective potential for each spin state from table \ref{D3tab}. All results are in Skyrme units with $\hbar=60$. }  \label{tab:C3wvs}
\end{table}

\begin{table}[ht]
	\centering
	\setlength\extrarowheight{5pt}
	\begin{tabular}{|l| c c c|}
		\hline
		&  \multicolumn{3}{c|}{Energy relative to ground state (MeV)}  \\ \hline
		State & Experiment & Calibration (i) & Calibration (ii) \\ \hline
		$\Ket{\Psi}_\frac{1}{2}$& $0.48$ & $-5.29$ & $-12.25$ \\ \hline
		$\Ket{\Psi}_\frac{3}{2}$& $0$ & $0$ & $0$ \\ \hline
		$\Ket{\Psi}_\frac{7}{2}$& $4.63$ & $4.14$ & $9.58$ \\ \hline
		$\Ket{\Psi}_\frac{5}{2}^{(1)}$& $6.68$ & $8.19$ & $18.97$ \\ \hline
		$\Ket{\Psi}_\frac{5}{2}^{(2)}$& $7.46$ & $18.79$ & $43.53$ \\ \hline			
	\end{tabular}
	\caption{A comparison of the experimentally obtained energy spectrum of $^7$Li (column 1) with the results from our calculation using Calibration (i) (column 2) and Calibration (ii) (column 3). The experimental data is from \cite{data}.} \label{Data}
\end{table}

The results are promising. All of the states considered are seen experimentally. The ordering is correct apart from the spin $\frac{1}{2}$ and $\frac{3}{2}$ states. We argued earlier that the spin $\frac{1}{2}$ state has higher energy than our calculation suggests as it must vanish in a subspace of $\mathcal{V}_5$. This may remedy the ordering issue. Most importantly, the spin $\frac{7}{2}$ state lies between the spin $\frac{3}{2}$ and $\frac{5}{2}$ states. The size of the gaps in the energy spectrum are reasonable for calibration (i) and much too large for calibration (ii). The ratios of the energy gaps between states are independent of this choice, though do depend on $\hbar$. We find that
\begin{equation}
\frac{E_{J=\frac{5}{2}} - E_{J=\frac{3}{2}}}{E_{J=\frac{7}{2}}-E_{J=\frac{3}{2}}}=1.98
\end{equation}
which is reasonably close to the experimental result, $1.44$.

The second spin $\frac{5}{2}$ state has very high energy. This can be understood by considering the body-fixed spin classically. The highly excited state has $|L_3|  = \frac{5}{2}$. This means that the spin is around the $3$-axis. This gives a large energy contribution since the Skyrmion is prolate in this direction. The lower energy spin $\frac{5}{2}$ state has $|L_3|  = \frac{1}{2}$ which allows it to rotate about an axis orthogonal to the prolate one. The state we have found probably does not correspond to the experimental state we have compared it to in table \ref{Data}. In the cluster model \cite{Tang} this state has a different structure than the others. It is described by a neutron orbiting a $^6$Li nucleus. Thus, it could be that we only see this spin state if we include a vibration which can split the $B=7$ Skyrmion into these clusters.

The next three experimental states of $^7$Li have spin $\frac{7}{2}$, $\frac{3}{2}$ and $\frac{3}{2}$. These have natural descriptions in this model. The $\frac{7}{2}$ state is orthogonal to the one allowed by the dodecahedron and has a single excited vibration in $\mathcal{V}_5$. The excited spin $\frac{3}{2}$ states will have isospin $\frac{3}{2}$, a possibility we neglected for simplicity. This would also describe the ground states of $^7$B and $^7$He which have spin $\frac{3}{2}$.

The mass of the $^7$Li nucleus is $6535$ MeV. Calibration (ii) gives the total mass of the ground state to be $6404$ MeV which is very close to the experimental value. Calibration (i) gives a much smaller value, only $2764$ MeV. There are several ways this could be remedied. First, we have only taken one of the Skyrmion's vibrational modes into account. There are approximately $6B$ modes, all of which will contribute to the energy. The Casimir energy contribution is also large, a $40\%$ correction in the $B=1$ sector \cite{Casimir}. Finally, the Lagrangian may be altered to include a $6$th order term which can be chosen to contribute positively to the mass. When this term is the same order as the other terms in the Lagrangian, the Skyrmion solutions do not change significantly \cite{6thTheses}. Thus the calculation in this paper would not vary greatly except for the total energy. These three factors could combine to give a reasonable value for the total mass. They also highlight the uncertainty in calculations of total mass in the Skyrme model.

\begin{figure}[ht]
	\centering
	\includegraphics[width=3in]{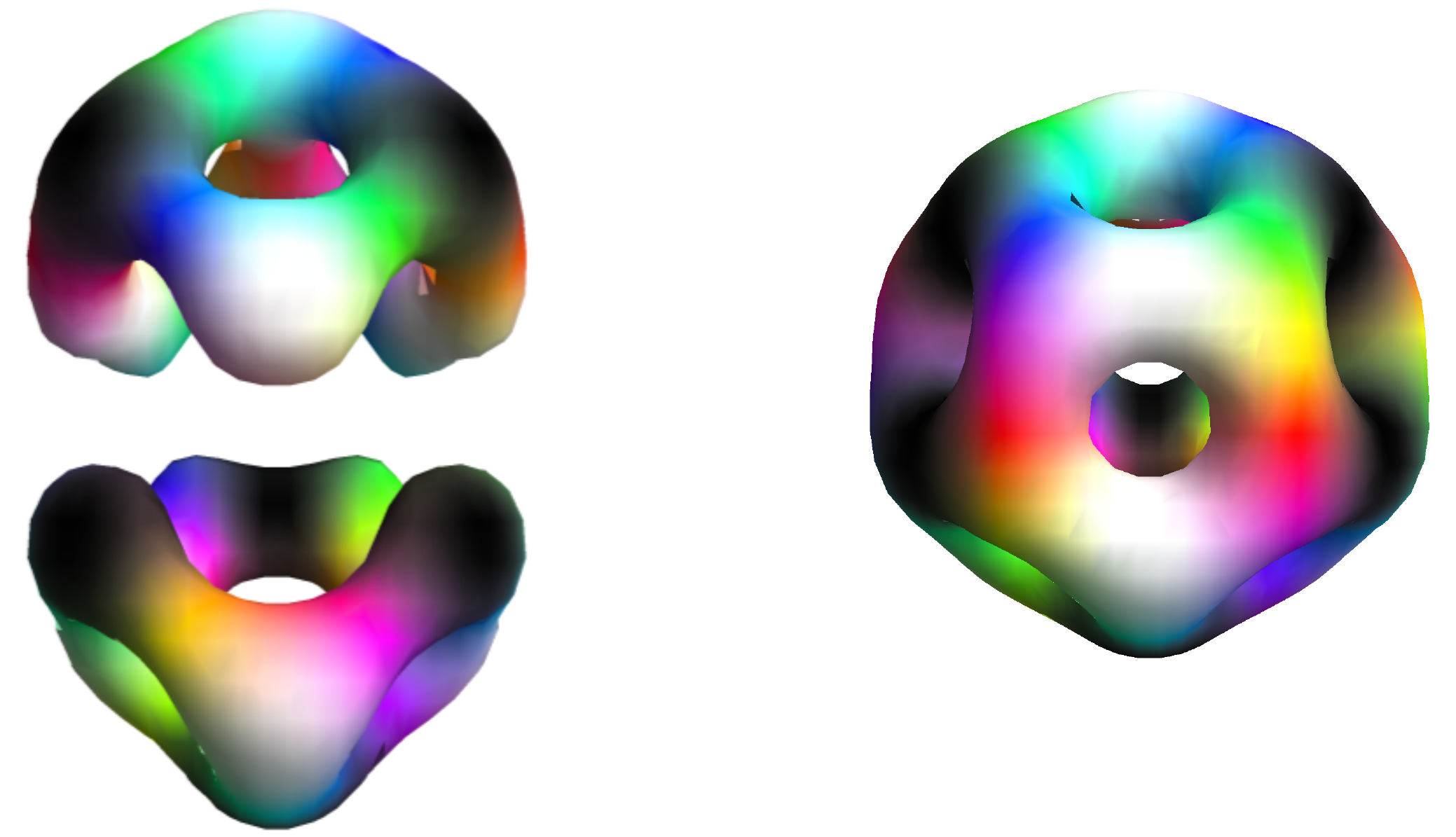}
	\caption{Plots of the baryon density at the maximum value of the vibrational wavefunctions. The spin $\frac{3}{2}$ state is on the left while the spin $\frac{7}{2}$ state is on the right.}
	\label{fig:baryondensity}
\end{figure}

Inclusion of the $C_3$ vibration has given us a good model of the spin $\frac{3}{2}$, $\frac{7}{2}$ and $\frac{5}{2}$ states of the $^7$Li/$^7$Be isodoublet. Further, it brings us closer to the cluster model of nuclei. This is apparent when we plot the classical baryon density at the maximum of the vibrational wavefunctions which shows us a classical approximation of the quantum state, before rotational averaging. These are plotted in figure \ref{fig:baryondensity}. We see that the spin $\frac{3}{2}$ state exhibits clustering while the spin $\frac{7}{2}$ state does not. This goes against conventional wisdom in the cluster model where the ground state is generally the most isotropic. 

\subsection{The $D_5$ direction}
	
To simplify this calculation we orient the Skyrmion as it is in figure \ref{fig:B7Sky}. The $D_5$ symmetry constrains $U$, $V$ and $W$ to be diagonal with
\begin{equation}
U_{11} = U_{22}, \, V_{11} = V_{22}, \, \text{and} \, W_{11} = W_{22} =0 \,  .
\end{equation}
The spin states allowed by $D_5$ symmetry are presented in table \ref{tab:D5spin}.  The spin $\frac{7}{2}$ state which is allowed by the dodecahedron is given by
\begin{equation}
\tilde{\ket{\Theta}}_{\frac{7}{2}}=\sqrt{\frac{7}{10}}\tilde{\ket{\Theta}}^{(1)}_{\frac{7}{2}} - \sqrt{\frac{3}{10}}\tilde{\ket{\Theta}}^{(2)}_{\frac{7}{2}} \, .
\end{equation}
	
\setlength\extrarowheight{6pt}
\begin{table}
	\begin{center}
		\begin{tabular}{|l | l|}
			\hline
			Spin & FR-allowed states \\ \hline
			$J=\frac{3}{2}$ & $\tilde{\ket{\Theta}}_{\frac{3}{2}} =\ket{\frac{3}{2},\frac{3}{2}}\ket{\frac{1}{2},-\frac{1}{2}}+\ket{\frac{3}{2},-\frac{3}{2}}\ket{\frac{1}{2},\frac{1}{2}}$ \\
			$J=\frac{5}{2}$ & $\tilde{\ket{\Theta}}_{\frac{5}{2}} =\ket{\frac{5}{2},\frac{3}{2}}\ket{\frac{1}{2},-\frac{1}{2}}-\ket{\frac{5}{2},-\frac{3}{2}}\ket{\frac{1}{2},\frac{1}{2}}$ \\
			$J=\frac{7}{2}$ & $\tilde{\ket{\Theta}}^{(1)}_{\frac{7}{2}} =\ket{\frac{7}{2},\frac{3}{2}}\ket{\frac{1}{2},-\frac{1}{2}}+\ket{\frac{7}{2},-\frac{3}{2}}\ket{\frac{1}{2},\frac{1}{2}}$ \\
			&
			$\tilde{\ket{\Theta}}^{(2)}_{\frac{7}{2}}=\ket{\frac{7}{2},\frac{7}{2}}\ket{\frac{1}{2},\frac{1}{2}}-\ket{\frac{7}{2},-\frac{7}{2}}\ket{\frac{1}{2},-\frac{1}{2}}$ \\ \hline
		\end{tabular}
	\end{center}
	\caption{The low energy states allowed by $D_5$ symmetry.} \label{tab:D5spin}
\end{table}

\begin{table}[b!]
	\centering
	\begin{tabular}{| l | c | l | l |} 
		\hline
		State & Vibrational wavefunction and potentials & $\mathcal{M}_7 + E_J(0) + \epsilon_\text{vib}$  & $E$ \\ \hline 
		$\tilde{\ket{\Psi}}_\frac{3}{2}$ & \includegraphics[width=2in]{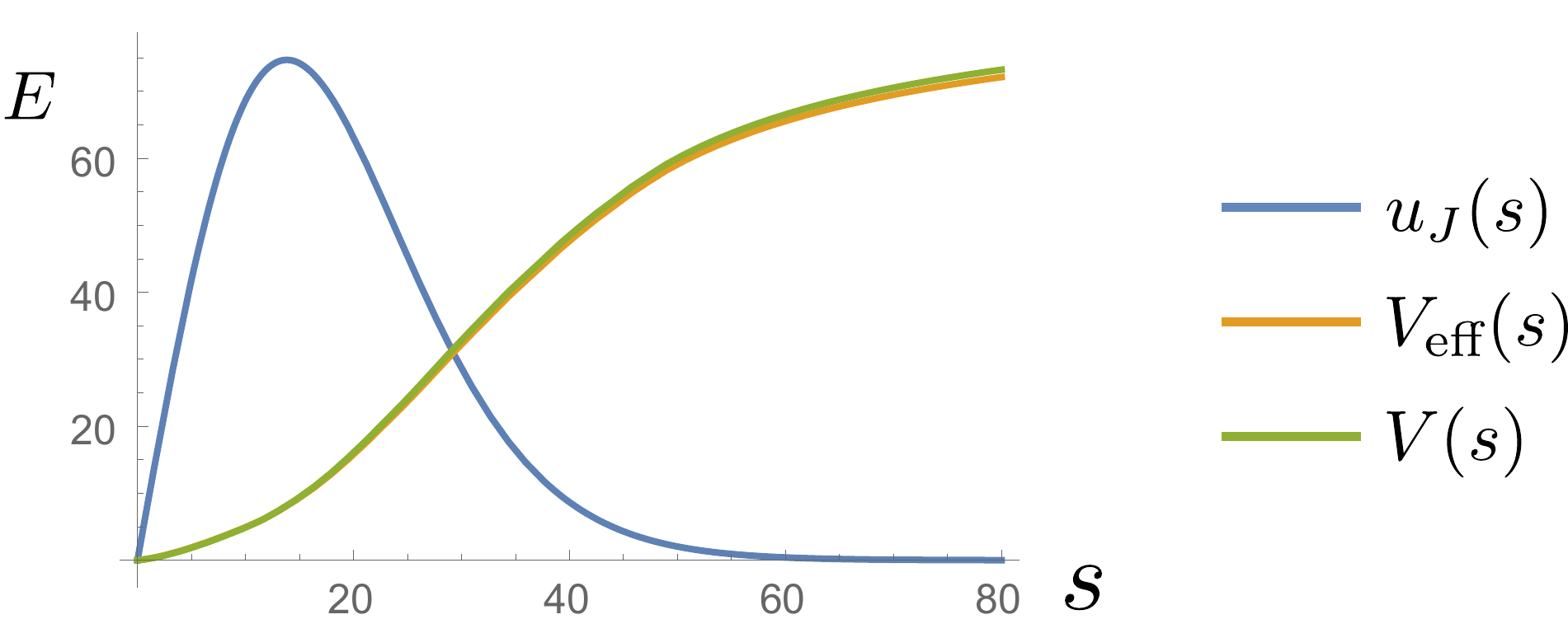} & 985.13 + 8.84 + 26.47 & 1020.45 \\ \hline
		$\tilde{\ket{\Psi}}_\frac{5}{2}$ & \includegraphics[width=2in]{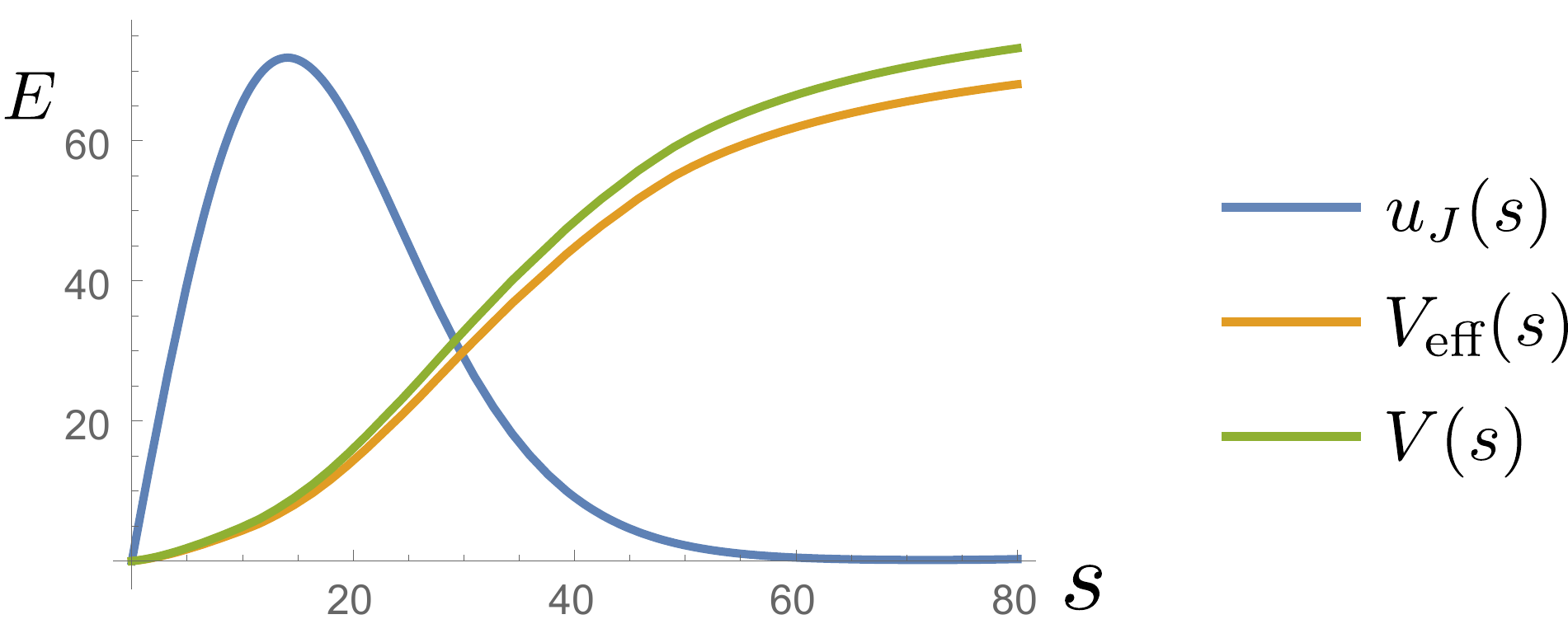} & 985.13 + 13.54 + 25.23 & 1023.91 \\ \hline
		$\tilde{\ket{\Psi}}_\frac{7}{2}$ & \includegraphics[width=2in]{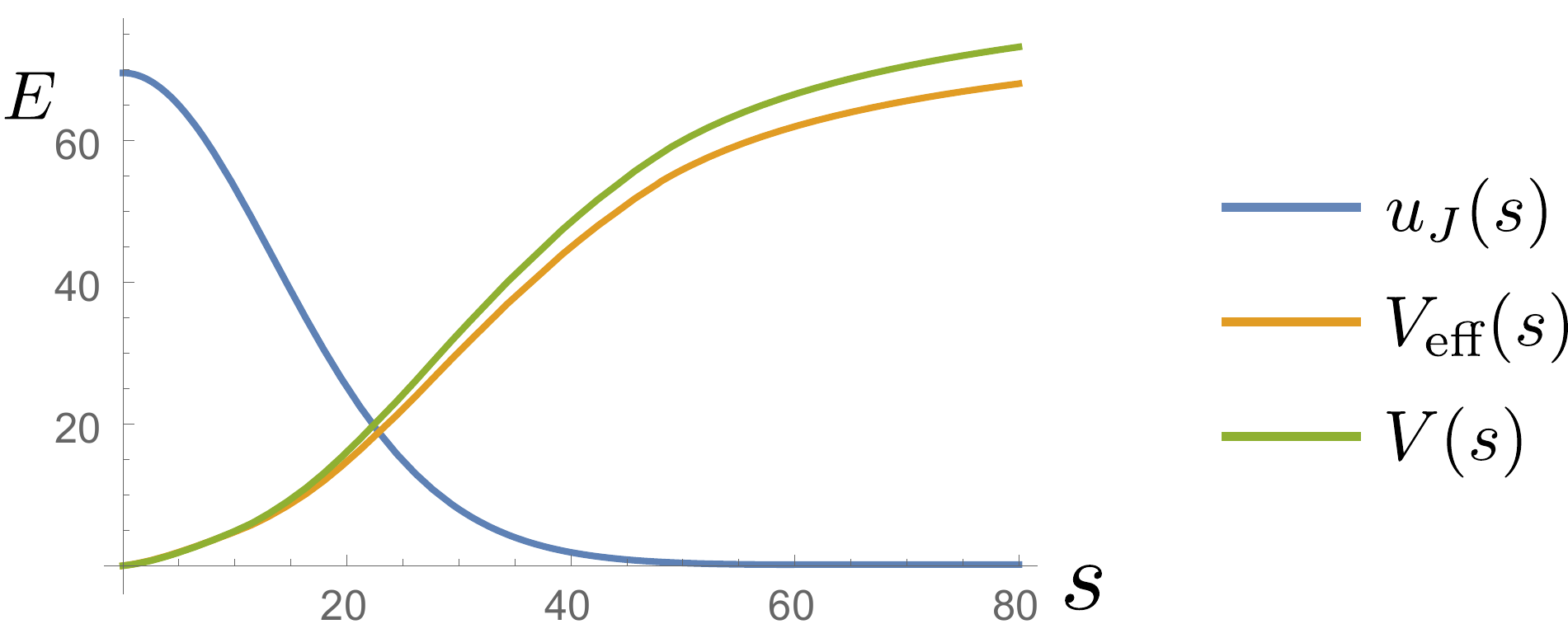} & 985.13 + 20.28 + 10.09 & 1015.50 \\ \hline
	\end{tabular}  
	\caption{The numerical results for quantisation along the $D_5$ direction. We display the vibrational wavefunction, potential and effective potential for each spin state from table \ref{tab:D5spin}. All results are in Skyrme units, with $\hbar=60$.  }\label{tab:D5}
\end{table}

We solve the Schr\"odinger equation and present the numerically generated vibrational wavefunction $u(s)$, potential $V(s)-\mathcal{M}_7$ and effective potential $V_\text{eff}(s)$ for each state in table \ref{tab:D5}. We also note the classical mass of the Skyrmion $\mathcal{M}_7$, the energy contribution from rotations $E_J(0)$ and the contribution from vibrations $\epsilon_\text{vib}$, as well as the total energy of each state $E$. We see that the $C_3$ direction produces lower energy states and should be considered the lower energy direction in $\mathcal{V}_5$. The states arising from the $D_5$ direction have higher energy than any experimentally discovered state and so are not relevant to the known energy spectrum of the $^7$Li/$^7$Be isodoublet. Earlier it was noted that the direction of the vibration in $\mathcal{V}_5$ is important. These results highlight this fact.
\clearpage
\subsection{The root mean square matter radius}

We saw in figure \ref{fig:baryondensity} that the different states appear to have different sizes. The simplest quantitative measure of the size of a nucleus is the root mean square (rms) matter radius, $\langle r_m\rangle$. We can calculate this for each value of $s$ by taking the square root of
\begin{equation}
r_m^2(s) = \frac{\int |\boldsymbol x|^2 \rho(\boldsymbol x,s) \, d^3  x}{\int  \rho(\boldsymbol x,s) \, d^3  x}
\end{equation}
where $\rho(\boldsymbol x,s)$ is the energy density of the Skyrme configuration at $s$. For a given state, the rms matter radius is then
\begin{equation}
\langle r_m\rangle = \bra{\Psi_J}r_m(s)\ket{\Psi_J} = \int r_m(s) u_J^2(s) \sqrt{|g|}\, ds
\end{equation}
where we have taken the vibrational wavefunctions to be normalised. We find that the matter radius of the spin $\frac{3}{2}$ state, in Skyrme units, is
\begin{equation} \label{rms}
\langle r_m\rangle_{\frac{3}{2}} = \, 1.85 .
\end{equation}
Experiments are unable to measure the matter radius directly. However in most nuclei the matter and charge radii are very similar. Thus we compare \eqref{rms} to the experimentally determined charge radius, $2.444$ fm. Our result depends on our choice of $F_\pi$. Calibration (i) gives a matter radius of $2.22$ fm, close to the experimental value. However calibration (ii) gives a very small radius, $0.96$ fm. Earlier we found that calibration (i) gave a better match to the energy spectrum. This result adds weight to the idea that it is the better choice. Regardless, ratios of lengths are independent of $F_\pi$. As such we can compare the matter radii for the spin $\frac{7}{2}$ and spin $\frac{3}{2}$ states and have more trust in the result. We find that
\begin{equation}
\frac{\langle r_m\rangle_{\frac{3}{2}}}{\langle r_m\rangle_{\frac{7}{2}}} = 1.07 \, .
\end{equation}
Thus we predict that the ground state of $^7$Li is $7 \%$ larger than the second excited state, which has spin $\frac{7}{2}$. The rms charge radius of an excited state is difficult to measure experimentally. As such there is no data to confirm our prediction. This is an important signature for the Skyrme model as this prediction is in conflict with the standard cluster model and shell model predictions.

\section{Conclusion and Outlook}
	
In this paper we have considered the inclusion of vibrational modes in the quantisation of the $B=7$ Skyrmion. We argued that to understand the low lying states of the $^7$Li/$^7$Be isodoublet one can truncate to quantisation along a $1$-dimensional line in vibrational space, $\mathcal{V}_5$. The space has a rich structure best understood using the geometry of a $5$-simplex. Using this, we picked special directions in the space to quantise along. The calculation gives a reasonable energy spectrum, much closer to the experimental data than had previously been found using zero mode quantisation. Most importantly, the spectrum includes all experimentally seen states and has the spin $\frac{7}{2}$ state lying above the spin $\frac{3}{2}$ state.

During the quantisation procedure some cluster structure emerged. We found that the Skyrmion picks out the $C_3$ direction as the lowest energy direction. This is remarkable since this is the vibration used in the basic $4+3$ cluster model. This brings the Skyrme model closer to the cluster models which are used widely in nuclear physics. The advantage of the Skyrme model is that the dynamics of the clusters are fully determined by the Lagrangian. They can merge smoothly into the $B=7$ Skyrmion or be infinitely separated; our formalism takes account of all configurations in between.
	
We predict that the excited spin $\frac{7}{2}$ state of $^7$Li is smaller than the spin $\frac{3}{2}$ ground state. The result depends crucially on the dodecahedral symmetry of the $B=7$ Skyrmion. This symmetry appears to persist in modified Skyrme models except in extreme BPS models \cite{lowbinding} \cite{BPSm}. Thus this prediction is an important signature for soliton models of finite nuclei.
	
Vibrational modes have the capacity to fix many issues in the Skyrme model including the high binding energies and small radii found using zero mode quantisation. They also have a fascinating and rich geometric structure. For these reasons alone, more work should be done to understand the vibrational spaces of Skyrmions. It is somewhat surprising that their inclusion leads to a resolution of problems in the $B=7$ sector. Hopefully a similar analysis in other sectors can produce more surprises.

\subsection*{Acknowledgments}

I would like to thank Professor Nick Manton for numerous useful discussions and for comments on the manuscript of this paper. I am grateful to Chris King for discussions regarding the structure of vibrational spaces. The work is supported by an STFC studentship.


\begin{thebibliography}{99}
		
		\bibitem{Sk} T.H.R. Skyrme, 
		A nonlinear field theory.
		{\it Proc. Roy. Soc. } {\bf A260} (1961) 127.
		
		\bibitem{AttDeu} R. Leese, N. S. Manton and B. Schroers,
		Attractive Channel Skyrmions and the Deuteron.
		{\it Nucl. Phys. } {\bf B442} (1995) 228.
		
		\bibitem{Light} O. Manko, N. S. Manton and S. Wood,
		Light nuclei as quantized Skyrmions.
		{\it Phys. Rev. }{\bf C76} (2007) 055203.
		
		\bibitem{Hoyle} P. H. C. Lau and N. S. Manton,
		States of Carbon-12 in the Skyrme Model.
		{\it Phys. Rev. Lett. }{\bf 113} (2014) 232503.
		
		\bibitem{Tang} Y. C. Tang, K. Wildermuth and L. D. Pearlstein,
		Cluster Model Calculation on the Energy Levels of the Lithium Isotopes.
		{\it Phys. Rev. }{\bf 123} (1961) 548.
		
		\bibitem{108} D. T. J. Feist, P. H. C. Lau and N. S. Manton,
		Skyrmions up to Baryon Number 108.
		{\it Phys. Rev. }{\bf D87} (2013) 085034.
		
		\bibitem{Kim} W. K. Baskerville,
		Vibrational spectrum of the $B=7$ Skyrme soliton.
		hep-th/9906063 (1999).
		
		\bibitem{LB} G. W. Gibbons and N. S. Manton,
		Classical and quantum dyanmics of BPS monopoles.
		{\it Nucl. Phys. }{\bf B274} (1986) 183.
		
		\bibitem{FRc} D. Finkelstein and J. Rubinstein,
		Connection between Spin, Statistics, and Kinks.
		{\it J. Math. Phys. }{\bf 9} (1968) 1762.
		
		\bibitem{RMan} C. J. Houghton, N. S. Manton and P. M. Sutcliffe,
		Rational maps, monopoles and Skyrmions.
		{\it Nucl. Phys. }{\bf B510} (1998) 507.
		
		\bibitem{Steffan} S. Krusch,
		Homotopy of Rational Maps and the Quantization of Skyrmions.
		{\it Ann. Phys. }{\bf 304} (2003) 103.
		
		\begin{comment}
		\bibitem{Olga7} O. V. Manko and N. S. Manton,
		On the spin of the $B=7$ Skyrmion.
		{\it J. Phys. }{\bf A40} (2007) 3683.
		\end{comment}
		
		\bibitem{Duet} E. Braaten and L. Carson,
		Deuteron as a Soliton in the Skyrme Model.
		{\it Phys. Rev. Lett. } {\bf 56} (1986) 3525.
		
		\bibitem{book} N. Manton and P. Sutcliffe, 
		Topological Solitons. 
		{\it Cambridge University Press: Cambridge} (2004).
		
		\bibitem{data} D. R. Tilley, C. M. Cheves, J. L. Godwin, G. M. Hale, H. M. Hofmann, J. H. Kelley, C. G. Sheu and H. R. Weller,
		Energy levels of light nuclei $A=5, 6, 7$.
		{\it Nucl. Phys. }{\bf A708} (2002) 3.
		
		\bibitem{Casimir} F. Meier and H. Walliser,
		Quantum Corrections to Baryon Properties in Chiral Soliton Models.
		{\it Phys. Rept. }{\bf 289} (1997) 383.
		
		\bibitem{6thTheses} I. Floratos,
		Multi-skyrmion solutions of a sixth order skyrme model.
		{\it Durham theses. } http://etheses.dur.ac.uk/3988 (2001).
		
		\bibitem{lowbinding} M. Gillard, D. Harland and J. M. Speight,
		Skyrmions with low binding energies.
		{\it Nucl. Phys. }{\bf B895} (2015) 272.
		
		\bibitem{BPSm} C. Adam, J. S\'anchez-Guill\'en and A. Wereszczy\'nski,
		A Skyrme-type proposal for baryonic matter.
		{\it Phys. Lett. }{\bf B691} (2010) 105.
		
		
	\end{thebibliography}
\end{document}